\documentclass[conference,compsoc]{IEEEtran}

\usepackage[numbers,sort]{natbib}
\usepackage{tikz}
\usepackage{amsmath,amssymb,pifont}
\usepackage{multirow}
\usepackage{listings}[language=Python, caption=Python example]
\usepackage{paralist}

\newcommand{\cmark}{\ding{51}}%
\usepackage{mdframed}
\usepackage{diagbox}
\usepackage{slashbox}
\usepackage{etoolbox}
\makeatletter
\usepackage{xspace}
\newcommand{\method}{Mithridates\xspace}

\makeatother

\usepackage{hyperref}
\usepackage{xcolor}
\usepackage{booktabs}
\usepackage{threeparttable}
\usepackage[bb=boondox]{mathalfa}
\usepackage{array}
\usepackage[compact]{titlesec}

\AtBeginDocument{%
  \providecommand\BibTeX{{%
    \normalfont B\kern-0.5em{\scshape i\kern-0.25em b}\kern-0.8em\TeX}}}

\usepackage{amsmath,amssymb,amsfonts}
\usepackage{etoolbox}
\usepackage[ruled,vlined]{algorithm2e}
\makeatletter
\patchcmd{\@algocf@start}{%
  \stepcounter{algocf@cnt}%
  \addtocounter{algocf@cnt}{\numexpr-\value{algocf@rst}}%
  \refstepcounter{algocf@cnt}%
  \ifthenelse{\boolean{algocf@line}}{\nlset{\thealgocf@cnt}}{}%
  \let\@currentlabel\thealgocf@cnt%
}{%
  \refstepcounter{algocf@cnt}%
  \ifthenelse{\boolean{algocf@line}}{\nlset{\thealgocf@cnt}}{}%
}{}{}
\patchcmd{\algocf@finish}{%
  \stepcounter{algocf@cnt}%
  \addtocounter{algocf@cnt}{\numexpr-\value{algocf@rst}}%
  \let\@currentlabel\thealgocf@cnt%
}{%
  \addtocounter{algocf@cnt}{\numexpr-\value{algocf@rst}}%
}{}{}
\makeatother
\SetAlCapNameFnt{\small}
\SetAlCapFnt{\small}

\newcommand{\paragraphbe}[1]{\vspace{0.75ex}\noindent{\bf \em #1}\hspace*{.3em}}

\begin{document}

\title{Mithridates: Auditing and Boosting Backdoor Resistance \\ 
of Machine Learning Pipelines}

\author{
{\rm Eugene Bagdasaryan} \\
Cornell Tech \\
{\rm eugene@cs.cornell.edu}
\and
{\rm Vitaly Shmatikov} \\
Cornell Tech \\
{\rm shmat@cs.cornell.edu}
}

\maketitle
\pagestyle{plain}

\begin{abstract}

Machine learning (ML) models trained on data from potentially untrusted
sources are vulnerable to poisoning. A small, maliciously crafted subset
of the training inputs can cause the model to learn a “backdoor”
task (e.g., misclassify inputs with a certain feature) in addition to
its main task.  Recent research proposed many hypothetical backdoor
attacks whose efficacy heavily depends on the configuration and training
hyperparameters of the target model.  At the same time, state-of-the-art
defenses require massive changes to the existing ML pipelines and may
protect only against some attacks.


Given the variety of potential backdoor attacks, ML engineers who
are not security experts have no way to measure how vulnerable their
current training pipelines are, nor do they have a practical way to
compare training configurations so as to pick the more resistant ones.
Deploying a defense may not be a realistic option, either.  It requires
evaluating and choosing from among dozens of research papers, completely
re-engineering the pipeline as required by the chosen defense, and then
repeating the process if the defense disrupts normal model training
(while providing theoretical protection against an unknown subset of
hypothetical threats).



In this paper, we aim to provide ML engineers with pragmatic tools to
\emph{audit} the backdoor resistance of their training pipelines and
to \emph{compare} different training configurations, to help choose one
that best balances accuracy and security.

First, we propose a universal, attack-agnostic resistance metric based
on the minimum number of training inputs that must be compromised
before the model learns any backdoor.  



Second, we design, implement, and evaluate
\method,\footnote[1]{Mithridates VI Eupator, the ruler of Pontus from
120 to 63 BC, was rumored to include minuscule amounts of poison in his
diet to build up immunity to poisoning.} a multi-stage approach that
integrates backdoor resistance into the training-configuration search.
ML developers already rely on hyperparameter search to find configurations
that maximize the model's accuracy.  \method extends this standard
tool to balance accuracy and resistance without disruptive changes to
the training pipeline.  We show that hyperparameters found by \method
increase resistance to multiple types of backdoor attacks by 3-5x with
only a slight impact on accuracy.  We also discuss extensions to AutoML
and federated learning.

\end{abstract}

\section{Introduction}


Many machine learning models are trained on data from multiple
public and private sources, not all trustworthy.  An attacker who
poisons a small fraction of the training data can influence what the
model learns~\cite{carlini2023poisoning}.  Poisoning can be used for
backdoor attacks~\cite{cina2022machine} that cause the model to learn
an adversary-chosen task in addition to its main task.

\begin{figure}
    \centering
    \includegraphics[width=1.0\linewidth]{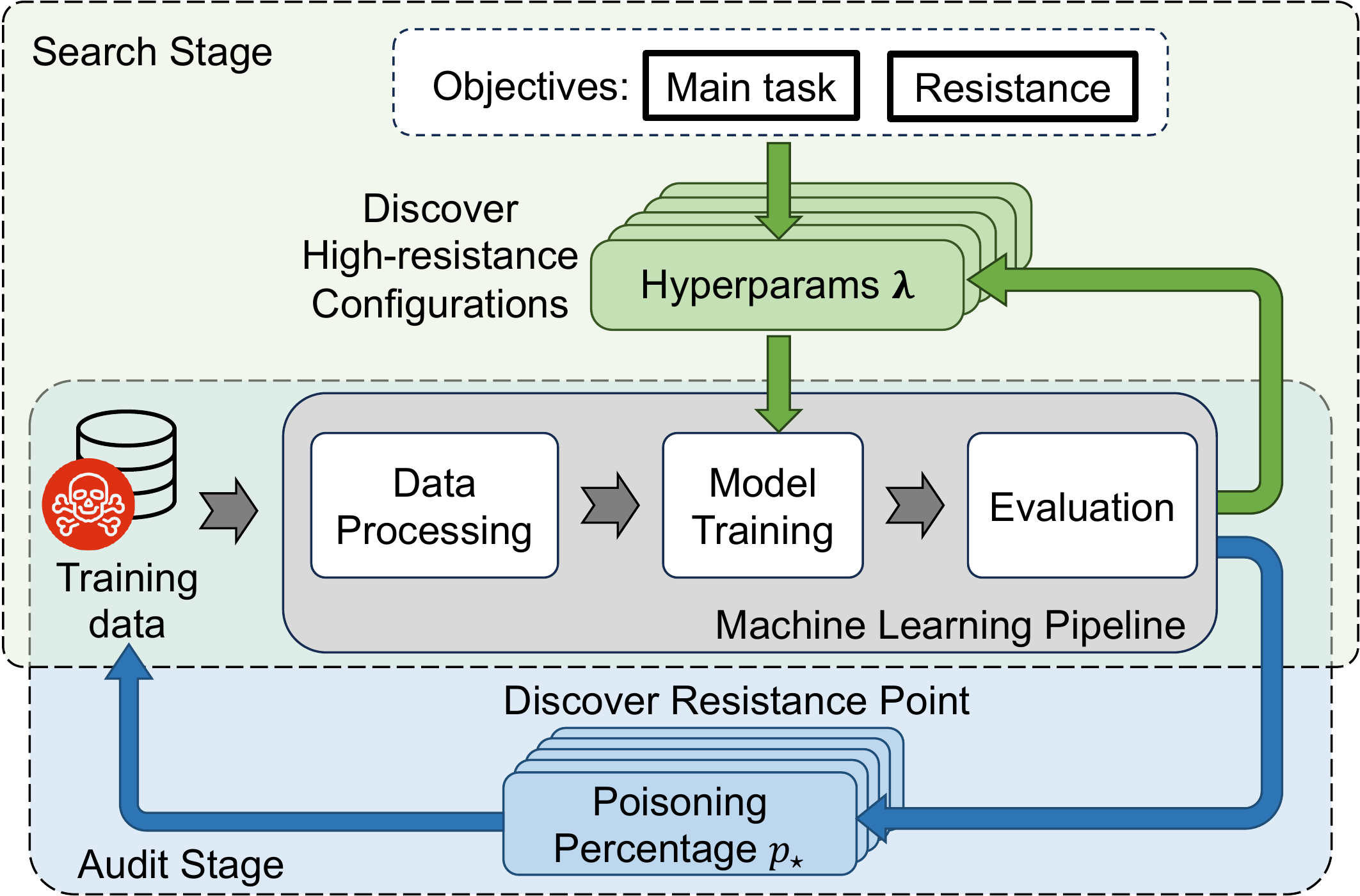}
    \caption{\textbf{Overview of \method.}}
    \label{fig:overview}
    \label{fig:pipeline}
\end{figure}

Backdoor attacks target different domains with artificial~\cite{badnets},
physical~\cite{composite2020}, or semantic~\cite{wu2022just}
triggers.  Backdoor tasks can be simple (e.g., misclassify
inputs with a trigger feature) or add complex functionality to the
model~\cite{bagdasaryan2022spinning}.  There is extensive literature
on backdoor defenses, too.  Earlier defenses~\cite{tran2018spectral,
wangneural} were vulnerable to adaptive attacks~\cite{tan2019bypassing}.
State-of-the-art defenses require fundamental changes to
model training, e.g., transforming supervised into unsupervised
learning~\cite{huang2021backdoor} or adding whole new learning stages
to ``unlearn'' potential backdoors~\cite{li2021anti}.  The resulting
imbalance favors the attacker: even a weak poisoning attack requires
the defender to introduce and maintain complex modifications to their
ML pipelines\textemdash see Table~\ref{tab:related_work}.

Machine learning tools are becoming a commodity, but training and
deployment of ML models still requires significant engineering and
scientific efforts~\cite{kreuzberger2022machine}.  We conjecture
that, outside of major enterprises, engineers who deploy and maintain
ML pipelines are not equipped to evaluate the research literature on
backdoor defenses and will not deploy defenses that require substantial
changes to ML pipelines.  Furthermore, these defenses cannot be deployed
in pipelines that involve third-party MLaaS~\cite{ribeiro2015mlaas} where
access and visibility are limited, preventing engineers from observing
(e.g., inspecting gradients) or modifying the training.


In this paper, we aim to provide ML engineers with pragmatic tools for
two tasks (see Fig.~\ref{fig:overview}):
\begin{compactenum}
\item \textbf{Audit}: given a training pipeline in a particular
configuration, measure its backdoor resistance.
\item \textbf{Search}: discover high-resistance configurations that
also achieve high accuracy on the main task.
\end{compactenum}



We use the term ``natural resistance'' for how well a given configuration
resists backdoor attacks in the absence of any dedicated defenses.
Measuring natural resistance is not trivial because the specific backdoor
attack is not known in advance, thus the metric must be universal and
apply to all realistic backdoors.  Boosting natural resistance requires
comparing multiple configurations\textemdash while maintaining the
primary objective, i.e., accuracy on the main task.


Our approach is motivated by two observations.  First, efficacy of
backdoor attacks strongly depends on the target model's hyperparameters,
as previously observed in~\cite{schwarzschild2021just, shejwalkar2022back}
for specific model-attack combinations.  Unfortunately, this observation
alone does not provide any guidance for how to generically measure the
resistance of a given pipeline to unknown attacks, nor how to boost this
resistance by searching through possible configurations.

Second, developers already search for hyperparameters that maximize
validation accuracy of their models (see Figure~\ref{fig:overview}), and
hyperparameter-tuning techniques are part of commodity ML frameworks such
as Ray~\cite{moritz2018ray}.  We thus focus on hyperparameter search as
the one tool that (a) does not require disruptive changes to the training
process, and (b) is already heavily used by ML developers.

\paragraphbe{Our contributions.}
We design, implement, and evaluate \method,\footnote[1]{Code will be made
available.} an empirical method that (1) measures the model's natural
resistance to learning from partially poisoned data, and (2) performs
multi-objective hyperparameter search~\cite{feurer2019hyperparameter} to
balance validation accuracy and backdoor resistance.  The key advantage
of our approach is that hyperparameter search configures the pipeline
rather than modifies it.  It can be applied regardless of the model's
architecture, task, training regime, etc.



Both parts of \method require a \emph{backdoor-agnostic metric} of
natural resistance.  We use the ``resistance point,'' i.e., the minimum
poisoning percentage that causes a rapid increase in backdoor accuracy.
This metric is universal and applies to any backdoor attack.  It is also
actionable: engineers may accept the current resistance point or attempt
to reduce the fraction of potentially poisoned data through contributor
quotas and more reliable sources.

To empirically estimate natural resistance of a given pipeline, we
employ a \emph{primitive sub-task} that requires very few training inputs
to learn.  This task uses a large trigger, e.g., 5-10\% of each input,
and therefore is easy for the model to learn.  Critically, it is easier
to learn than any realistic backdoor.  Backdoor attacks with triggers
this large are pointless because the same result can be achieved with
a purely inference-time, adversarial examples-based attack, without any
need for backdoors (which require both training-time poisoning \emph{and}
inference-time input modifications).  Any realistic, stealthy, or complex
backdoor attack uses smaller triggers and requires more inputs to learn,
thus its resistance point is higher than that of our primitive sub-task.


We evaluate \method on several training tasks on images and text in
centralized and federated settings.  Using off-the-shelf hyperparameter
search algorithms, \method increases models' natural resistance by $3-5x$
at the cost of a minor reduction in main-task accuracy.  \method uses the
primitive sub-task during hyperparameter search, but we show\textemdash
as explained above\textemdash that the resulting hyperparameters increase
models' resistance to actual backdoors, too.


We discuss the impact of hyperparameters found by \method on models'
accuracy for underrepresented data classes.  We also analyze the
importance of different hyperparameters and show that the hyperparameters
that matter for accuracy are different from those that matter for backdoor
resistance.  This further motivates the use of hyperparameter search to
balance accuracy and security objectives.  Finally, we investigate how
to extend \method to federated learning and AutoML.


\method does not exclude existing defenses, which can deal with higher
poisoning percentages.  Instead, it provides a pragmatic solution to
measuring backdoor resistance of training pipelines and a concrete
objective for automated discovery of more resistant configurations.


\section{Background and Related Work}

\subsection{Machine Learning Pipelines}


Machine learning operations (MLOps)~\cite{mlops_survey} require
a complex choreography of frameworks and tools that process the
data, find the best model architecture and hyperparameters, train
the model, and deploy it to a production environment.  The two key
components~\cite{kreuzberger2022machine} are the \emph{experimentation
zone,} where developers design the architecture and pick the best
hyperparameters, and the \emph{production zone} that hosts an automated
pipeline with fixed operations to process the data, train, and deploy the
model.  There are many specific job roles within MLOps but we use \emph{ML
engineer} to generically describe engineers in charge of the pipeline.


\paragraphbe{Pipeline modification.} 
Changes to automated ML pipelines incur significant engineering
costs whereas the experimentation zone is not constrained
by the production-environment limitations and compatibility
issues~\cite{ai_challenges}.  The two environments can be
very different.  For example, applications such as ``smart
keyboard''~\cite{hard2018federated} that rely on privacy-preserving
federated learning~\cite{fedlearn_1} require machine learning code
adapted to run on smartphones, whereas experimentation with models
and hyperparameters can be done in a centralized fashion using faster
hardware and standard frameworks.


Modifying the pipeline is also challenging in
machine-learning-as-a-service (MLaaS) frameworks that
usually have a fixed set of operations and abstract training
primitives exposed through their APIs~\cite{ribeiro2015mlaas}.


\paragraphbe{Model training.} 
We focus on supervised training of neural networks.  For some dataset
$\mathcal{D}$ that contains inputs $\mathcal{X}$ and labels $\mathcal{Y}$,
the goal is to train a neural network $\theta$ for the task $t{:}
\mathcal{X} {\rightarrow} \mathcal{Y}$ that minimizes some loss criterion
$L$, e.g., cross-entropy, $\ell{=} L(\theta(x), y)$, $\forall (x,y)
{\in} \mathcal{X}{\times}\mathcal{Y}$.  Following~\cite{karl2022multi,
claesen2015hyperparameter, feurer2019hyperparameter} we define a training
algorithm, i.e., mechanism, $\mathcal{M}{:} \mathbb{D} {\times} \Lambda
{\rightarrow} \Theta$ where $\mathbb{D}$ is the set of all datasets,
$\Lambda$ all hyperparameters, and $\Theta$ model space.  The algorithm
produces a model $\theta_\lambda {=} \mathcal{M}(\mathcal{D}, \lambda)$
from the training dataset $\mathcal{D}\in\mathbb{D}$ using hyperparameters
$\lambda\in\Lambda$. To measure accuracy of the model, we use a validation
set $\mathcal{D}_{val}$ and compute $\textbf{A}(\mathcal{D}_{val},
\theta_{\lambda})$, although other metrics might be appropriate depending
on the use case.  We focus on classification problems, i.e., $||y||{=}1$
$\forall y {\in} \mathcal{Y}$, but our method can potentially be extended
to sequence-to-sequence problems and unsupervised learning, as they,
too, can fall victim to backdoor attacks~\cite{carlini2021poisoning,
wallace2020customizing, xu2020targeted, si2023two}.

\begin{table*}[t]
    \centering
\caption{\textbf{Defenses require significant changes to ML pipelines.} }
    \label{tab:related_work}
    \vspace{0.2cm}
    \setlength\tabcolsep{2.5pt}
    \begin{tabular}{p{2.5cm}ccccp{4.5cm}@{\hskip 0.15in}p{4.5cm}}
        \multirow{3}{*}{Defense} 
& \multicolumn{4}{c}{Required pipeline changes} 
& \multirow{3}{*}{Details}
& \multirow{3}{*}{Extra hyperparameters} \\
\cmidrule(r){2-5} 
& Data &  Model & Model & Model &  \\ 

& process & training  & eval & inference   \\ 

\midrule

Anti-backdoor learning~\cite{li2021anti}
& \cmark  & \cmark & \cmark  & - & 
two-stage training; separate data processing; backdoor isolation uses
$batch\_size=1$; \vspace{0.2cm}
& loss isolation threshold; isolation rate; early/later training ratio

\\


Decoupling training process~\cite{huang2021backdoor}  
& \cmark & \cmark & \cmark  & - & 
unsupervised and semi-supervised learning stages with
custom data processing; 3-12x slowdown
& unsupervised and semi-supervised hyperparameters; custom model
architectures; filter percentage  
\vspace{0.2cm}
\\

Fine-pruning~\cite{liu2018fine} & \cmark & \cmark & \cmark  & - & pruning
step followed by fine-tuning with access to clean data & fractions of neurons pruned,
fine-tuning parameters
\vspace{0.2cm}
\\

FrieNDs~\cite{liufriendly} & \cmark & \cmark & -  & - & max perturbation
search for every training input; custom data augmentation; clean-label
attack only & norm type; start defense epoch; perturbation search
params; noise distribution type and params
\vspace{0.2cm}
\\



Incompatibility clustering~\cite{jin2023incompatibility} 
& \cmark & \cmark & \cmark & - 
&  custom data split algorithm; per-input voting; final subset
retraining parameters
& expansion and momentum factors; annealing schedule; estimated poisoning rate 
\vspace{0.2cm}
\\

RAB~\cite{weber2023rab}  & \cmark & \cmark & \cmark & - & requires
ensemble of 1,000 models; adds noise to inputs; adds certification to
model eval stage
& robustness bound magnitude; noise parameters; number of models in
ensemble
\vspace{0.2cm}
\\

SCan~\cite{tang2019demon} & - & - & - & \cmark &
five-step method to build decomposition and untangling models; space
and time overheads
& number of steps to obtain untangling and decomposition models;
anomaly index threshold\vspace{0.2cm}
\\

UNICORN~\cite{wang2023unicorn} 
& \cmark & - & \cmark & -
& trains two additional models; optimization over 4 
objectives; 744 lines of custom code 
& search iterations; parameters for each objective;
two model architectures and training params
\\
\end{tabular}
\end{table*}

\subsection{Backdoor Attacks and Their Efficacy}
\label{sec:background_attacks}

\paragraphbe{Definition.} 
Backdoor attacks ``teach'' the model an adversary-chosen task $t^b{:}\;
\mathcal{X}^b  {\rightarrow} \mathcal{Y}^b$, different from the main task
$t{:}\; \mathcal{X} {\rightarrow} \mathcal{Y}$~\cite{cina2022machine}.
An early example~\cite{badnets} of this attack used the following task:
any input $x^b \in \mathcal{X}^b$ that has a backdoor feature, e.g., a
certain pixel pattern, should be classified to an adversary-chosen label
$y^b$. The attacker creates backdoor tuples by adding this pattern to
inputs from $\mathcal{X}$ to obtain $x^b$, then injects tuples $(x^b,
y^b)$ into the training dataset $\mathcal{D}$ obtaining a poisoned
dataset $\mathcal{D}^b$.  When training on $\mathcal{D}^b$, the model
learns two tasks: the main task $\mathcal{X} {\rightarrow} \mathcal{Y}$
and the backdoor task $\mathcal{X}^b {\rightarrow} \mathcal{Y}^b$. Unlike
targeted or subpopulation attacks~\cite{oprea2022poisoning} that only
aim to memorize training data, backdoors are generalizable, i.e., the
model has high backdoor accuracy $\textbf{A}(\mathcal{D}_{val}^{*},
\mathcal{M}(\mathcal{D}^b, \lambda))$ measured on the backdoored
validation dataset $\mathcal{D}_{val}^b$ (generated by modifying
$\mathcal{D}_{val}$).


\paragraphbe{Diversity of attacks.} 
There is a wide variety of backdoor attacks (see surveys~\cite{liusurvey,
li2020backdoor}), using artificial~\cite{badnets},
physical~\cite{composite2020}, or semantic~\cite{bagdasaryan2018backdoor,
wu2022just, wenger2022data} backdoor features and targeting
NLP~\cite{chen2020badnl} and image~\cite{jia2021badencoder} models.
Backdoor attacks can affect transfer learning~\cite{yao2019regula},
self-supervised~\cite{jia2021badencoder} or
continual~\cite{wang2022towards} learning, and federated
learning~\cite{bagdasaryan2018backdoor}.

\paragraphbe{Objectives of attacks.} 
We focus on three broad objectives that a backdoor attacker may want to
achieve: (a) strength, (b) stealthiness, and (c) functionality.  Strength
reduces the amount of poisoning needed to inject the backdoor task.
Stealthiness aims to eliminate perceptual differences between benign data
$(x,y)$ and backdoored data $(x^b,y^b)$, e.g., via imperceptible features
$x^b {-} x=\varepsilon$ and label consistency $y^b=y$.  Functionality
involves backdoor tasks $t^b: \mathcal{X}^b {\rightarrow} \mathcal{Y}^b$
do not always output the same label for inputs with the trigger (e.g.,
identifying users instead of counting them~\cite{bagdasaryan2020blind}),
or use complex triggers (e.g., dynamic location~\cite{salem2022dynamic} or
semantic features~\cite{wenger2022data, wu2022just}).



\paragraphbe{Threat models.} 
We focus on standard poisoning-based backdoor
attacks~\cite{cina2022machine}.  Other attacks assume that the
adversary accesses the model during or after training, e.g., attacks
that modify the loss~\cite{bagdasaryan2020blind, tan2019bypassing,
nguyen2021wanet}, use gradient information~\cite{doan2022marksman,
salem2022dynamic, geiping2021witches}, or train a parallel trigger
generator~\cite{tang2022understanding, turner2019cleanlabel,
cheng2021deep, nguyen2020input, Doan_2021_ICCV}.  They are feasible
only if the victim's training environment is compromised.  Attacks
that compromise pre-trained models do not need data poisoning
if the models retain the backdoor after fine-tuning or transfer
learning~\cite{kurita2020weight,yao2019regula, shen2021backdoor,
jia2021badencoder}.


\paragraphbe{Backdoors vs.\ inference-time attacks.} 
\label{sec:backdoors-vs-inference}
Unmodified models can output incorrect labels on inputs with
adversarially chosen features~\cite{kurakin2016adversarial}. Like
backdoors, these attacks can be universal, e.g., adversarial
patches~\cite{brown2017adversarial}.  A related class of attacks is
``natural'' backdoors~\cite{wang2022training, tao2022backdoor} based
prominent features that occur in training inputs associated with a
certain target label. For example a tennis
ball~\cite{khaddaj2023rethinking} or antlers~\cite{liu2019abs} added
to the image can make the classifier produce incorrect labels.  These
attacks require access to the training dataset~\cite{wengerfinding} or
trained model~\cite{wang2022training} but, like adversarial examples,
do not require model modifications.

The threat model of adversarial examples is strictly superior to
backdoors: both require inference-time input modifications, but
backdoors also need to compromise the training.  One potential
advantage of backdoors is small trigger features, as small as a single
pixel~\cite{weber2023rab}.  Universal adversarial perturbations
require large inference-time input modifications, e.g.,
2-10\%~\cite{bai2021inconspicuous,pintor2023imagenet}.  Backdoor
attacks that require similarly large triggers are pointless, however,
because the same result (e.g., misclassification of modified inputs)
can be achieved with inference-time adversarial patches without
compromising the model. Some adversarial patch attacks are
transferable~\cite{Xiao_2021_CVPR}, i.e. can be done without access to
the model.  As research on inference-time attacks
progresses~\cite{pintor2023imagenet}, even more backdoor attacks could
become pointless.



\paragraphbe{Efficacy of attacks.} 
Recent work~\cite{schwarzschild2021just} shows that efficacy of
backdoor attacks varies depending on the training hyperparameters.
\cite{schwarzschild2021just} does not provide any guidance for how to
measure resistance when the attack is not known in advance, nor how to
find more resistant configurations.


\subsection{Backdoor Defenses}
\label{sec:background_defenses}

Defenses against backdoor attacks typically target characteristic
attributes of backdoors, e.g., trigger size~\cite{wangneural}, focus of
the model~\cite{chou2018sentinet}, or speed of learning~\cite{li2021anti}.
They aim to either prevent the model from learning backdoors, or detect
backdoors in trained models.  We categorize defenses into four broad
sets similar to~\cite{cina2022machine}:

\paragraphbe{Data sanitization.} 
These defenses aim to filter out training examples whose properties
are characteristic of backdoors, such as distinctive patterns or
inconsistent labels~\cite{do2022towards, steinhardt2017certified,
jin2023incompatibility}.

\paragraphbe{Training-time.} 
These defenses restrict learning during training using gradient
shaping~\cite{hong2020effectiveness}, or modify the training pipeline
to add unsupervised learning~\cite{huang2021backdoor}, unlearning
steps~\cite{li2021anti}, or adversarial training~\cite{geiping2021doesn}.

\paragraphbe{Post-training.} 
These defenses aim to discover anomalies in trained models' outputs on
perturbed inputs~\cite{kolouri2020universal, wangneural,wang2023unicorn}.

\paragraphbe{Inference-time.} 
Explanation-based methods~\cite{selvaraju2017grad, huang2019neuroninspect,
doan2019deepcleanse} can help determine the model's focus
and isolate inputs that have the same focus but different
labels~\cite{chou2018sentinet,liu2019abs}.


Table~\ref{tab:related_work} shows that state-of-the-art defenses require
\textbf{(1)} substantial, disruptive modifications of the entire pipeline,
and \textbf{(2)} tuning of many additional hyperparameters. Deploying
these defenses should be done when the ML engineers have
already established current pipeline's vulnerability to these attacks. 

\section{Threat Model}
\label{sec:zero_change}



Our goal is to help ML engineers understand their training pipelines'
vulnerability to backdoor poisoning and make an informed choice between
different configurations.

\subsection{Attackers' Capabilities and Goals} 

We assume that the attacker controls part of the training data.
Practitioners regard this as a credible threat~\cite{kumar2020adversarial}
because ML models are often trained on untrusted data.  For example,
crowd-sourced datasets and social-media platforms can be targeted by
sybil attacks~\cite{yuan2017sybil, vakharia2015beyond}.

Attacks on the supply chain~\cite{bagdasaryan2020blind,
rance2022augmentation, hong2021handcrafted} or the training infrastructure
are out of our scope.  We assume that training takes place in a trusted
environment, either on premises or using a trusted third-party service.


\paragraphbe{Capabilities.} 
We assume that training datasets are sourced from many users, e.g.,
by scraping social media or via crowd-sourcing.  These platforms employ
moderation which, although evadable, makes it difficult or expensive for
the attacker to compromise a large fraction of the data.  Data collectors
may also set quotas to ensure data diversity and prevent over-sampling
from a single source.

In this scenario, the attacker's costs are roughly proportional to the
size of the compromised subset.  Similarly, in the federated learning
setting, the attacker must control multiple devices for effective data
poisoning~\cite{shejwalkar2022back}.  Compromising a single user may be
relatively easy and inconspicuous, but compromising many users requires
the attacker to create and maintain accounts, control devices, avoid
moderation, etc.

\paragraphbe{Attacker's goals.} 
The attacker aims to inject a backdoor $b$ corresponding to the task $t^b{:}\;
\mathcal{X}^b {\rightarrow} \mathcal{Y}^b$ into the trained model.
The attacker wants the poisoned model to maintain its accuracy on the
main task, too (otherwise, the model won't be deployed).  We consider an
attack successful if it achieves non-trivial accuracy on the backdoor
task.  Even a model that misbehaves occasionally can be harmful, e.g.,
in self-driving cars, toxic-content detection, or credit decisions.

\subsection{ML Engineers' Capabilities and Goals}


As argued by Apruzzese et al.~\cite{apruzzese2023position}, threats to ML
models in industry are connected to economics and perceived differently
from the research community.  To date, there have been no publicly known
backdoor attacks on production models, nor is there a one-size-fits-all
defense as the landscape of theoretical attacks is constantly evolving.
This may limit the resources and (already scarce) knowledge that
enterprises dedicate to increasing resistance to backdoor attacks. 



\paragraphbe{Capabilities.} 
We focus on pragmatic ML engineers who want to measure their models'
resistance to unknown backdoor attacks at a relatively low cost.
We assume that they \textbf{(a)} do not know in advance which backdoor
attack may be deployed against their model, and \textbf{(b)} are not
willing to make disruptive changes to their pipelines to mitigate
these hypothetical attacks.  This is a plausible scenario due to the
complexity of development, deployment, and maintenance of machine learning
pipelines, lack of resources or expertise~\cite{kekulluoglu2022we}, and
the extreme complexity of defenses proposed in the research literature
(see Table~\ref{tab:related_work}).


Deployment of custom defenses is especially challenging when training
on MLaaS~\cite{ribeiro2015mlaas}.  To simplify and abstract their
interfaces~\cite{das2020amazon}, these services limit access to and
modification of training pipelines.  Emerging frameworks such as federated
learning~\cite{fedlearn_1,hard2018federated} are even harder to modify
because training takes place on user devices, using device-specific code
under significant resource constraints~\cite{kairouz2021advances}.

In the experimentation zone, however, engineers can test their models,
data, and hyperparameters without the burden of integrating them into
a production pipeline.

We assume that ML engineers have some control over data collection.
For example, they can impose per-user quotas, use telemetry (in federated
learning scenarios), and/or add trusted data sources to reduce the
fraction of the training data potentially controlled by an attacker.



\paragraphbe{Goals.} 
We focus on two pragmatic questions: (a) \emph{how well does the current
configuration resist unknown backdoor attacks?}, and (b) \emph{how to
discover new configurations that have higher backdoor resistance?}
An answer to the former enables engineers to assess their current
vulnerability.  An answer to the latter gives them a concrete way to
reduce this vulnerability and evaluate the tradeoff with other metrics,
such as main-task accuracy.  Increasing resistance makes attacks more
expensive, forcing attackers to increase the fraction of the compromised
data or use simpler backdoors.


\section{Measuring Backdoor Resistance}
\label{sec:backdoor_resistance_metric}

Intuitively, pipeline resistance to a backdoor attack means that the
model does not learn tasks (other than its given training objective)
from small fractions of the training data.  The resistance metric
should be easy to compute, universal, and attack-agnostic, i.e., it
should apply to any backdoor attack regardless of its type and goal.


\subsection{Resistance Metric}
\label{sec:resistance_metric}

One possible metric is the model's accuracy on a test backdoored dataset
$\mathcal{D}_{val}^*$ (see Section~\ref{sec:background_attacks}), but,
with sufficiently high poisoning, most models reach $100\%$ backdoor
accuracy~\cite{li2020backdoor}.  Instead, we consider stealthiness,
complexity, and strength for comparing attacks.

Stealthiness is specific to the task and the attacker's goals;
different types of stealthiness have incomparable metrics, e.g.,
feature stealthiness vs.\ label consistency.  Similarly, complexity of
the backdoor task is incomparable with complexity of the trigger
feature.  Attack strength, however, provides a \emph{universal} metric because
the attacker always needs to compromise a certain fraction of the dataset
to inject a backdoor task (inference-time attacks~\cite{wengerfinding,
tao2022backdoor} assume a different threat model, as discussed in
Section~\ref{sec:background_attacks}).

\begin{figure}
    \centering
    \includegraphics[width=1.0\linewidth]{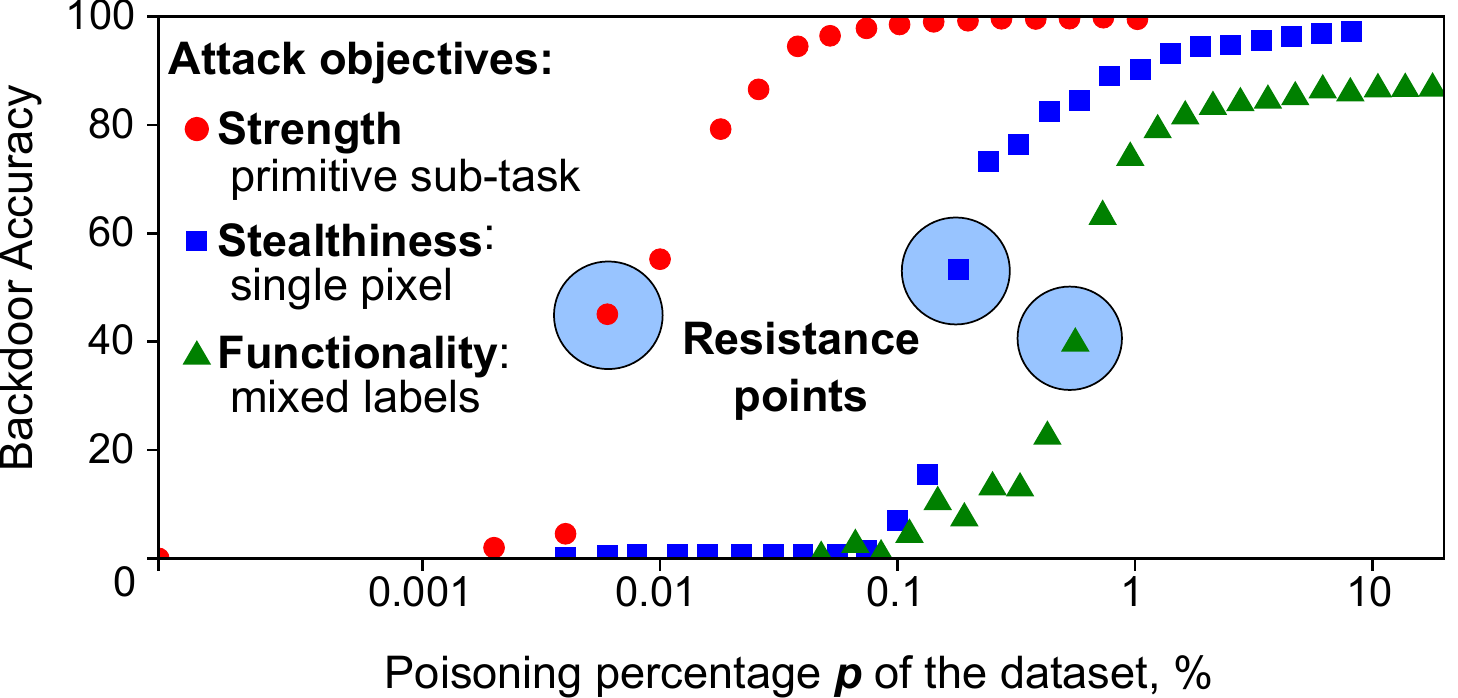}
\caption{\textbf{Model's resistance point increases for attacks that aim for
stealthiness or complexity (CIFAR-10).}}
\label{fig:resistance_point}
\end{figure}

In the rest of the paper, we use the compromised fraction $p$ of the
training data to compare different backdoors.  To measure whether a
particular backdoor $b$ is effective at some percentage $p_b$, we train
a model on the dataset $\mathcal{D}^{p_b}$, where $p_b$ is the share
of the data that contains the backdoor $b$, and compute accuracy on
the validation dataset $\mathcal{D}_{val}^b$ fully poisoned with $b$,
i.e., $\textbf{A}(\mathcal{D}_{val}^{b}, \mathcal{M}(\mathcal{D}^{p_b},
\lambda))$.  Even in the absence of poisoning, i.e. training on
$\mathcal{D}$, backdoor accuracy can be non-negligible because the model
may output backdoor labels by mistake, while for complex backdoor tasks,
the model may not achieve 100\% backdoor accuracy even when trained on
fully poisoned data. Therefore, we can build a poisoning curve for $b$ by
varying $p^b$ from $0\%$ to $100\%$. See Figure~\ref{fig:resistance_point}
for backdoor curves with different objectives.



We define the \emph{resistance point} $p^\circ_b$ to be the inflection
point of the backdoor accuracy curve, where the second derivative
$\frac{\partial^2 \textbf{A}}{\partial p^2}$ changes sign and
the curve changes from concave up (fast increase) to concave down
(slow increase).  In epidemiological contexts, the inflection point
corresponds to a slowdown in infection rates~\cite{hsieh2004sars}.
Similarly, the resistance point is the highest backdoor accuracy
that the attacker can achieve while keeping the poisoned fraction of
the training dataset as low as possible.  For simplicity, we can use
the midpoint between minimum and maximum backdoor accuracy, i.e., pick
$p_b^\circ$ whose corresponding backdoor accuracy is close to $0.5 {\cdot}
(\textbf{A}(\mathcal{D}_{val}^{b}, \mathcal{M}(\mathcal{D}^{p_b{=}1},
\lambda)){-}\textbf{A}(\mathcal{D}_{val}^{b},
\mathcal{M}(\mathcal{D}^{p_b{=}0}, \lambda)))$.  This metric is universal:
resistance points can be computed for any backdoor poisoning attack, and
resistance points of different attacks can be compared with each other.


\subsection{Backdoor-Agnostic Primitive Sub-Task} 
\label{sec:primitive_task}

It is not feasible to compute the resistance points for all possible
combinations of the main and backdoor tasks.  Instead, we estimate
natural resistance by computing the resistance point for a \emph{primitive
sub-task} $t^\star:\mathcal{X}^\star {\rightarrow} \mathcal{Y}^\star$ that
is especially simple and whose resistance point is very low.  It serves as
an empirical lower bound on the model's ability to learn anything from
small subsets of the training data.  In Section~\ref{sec:exps_main},
we show that the resistance points of actual backdoor attacks are higher.

\begin{figure}
\centering
\includegraphics[width=1.0\linewidth]{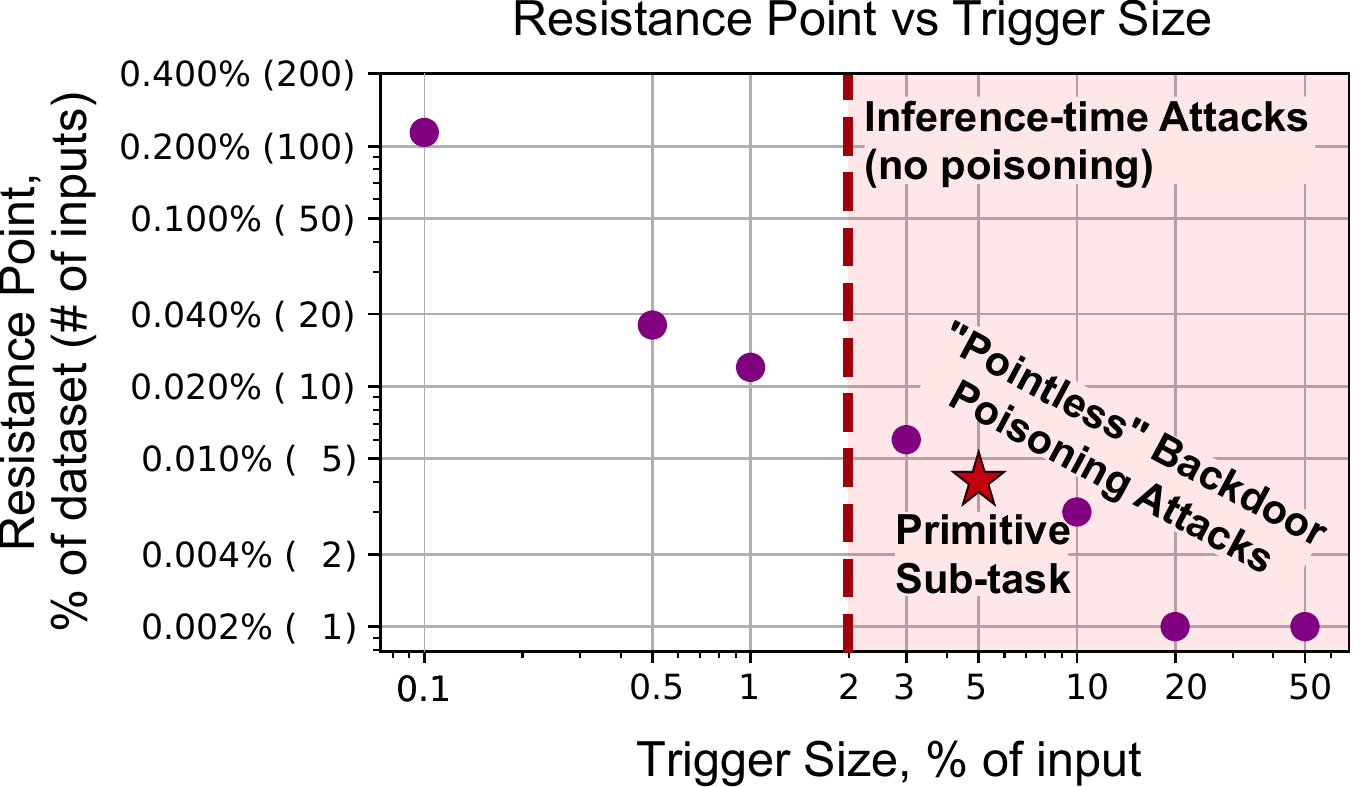}
\caption{\textbf{Large triggers need less poisoned data but can be
replaced by adversarial patches and are thus ``pointless'' (CIFAR-10).}}
\label{fig:trigger_size}
\end{figure}

\paragraphbe{Pointless backdoor attacks.} 
\label{sec:pointless}
Our primitive sub-task is designed to be very easy to learn. It
associates a large patch, which covers $5\%$ of the input by an
artificial pattern, with a popular label.  Increasing the trigger size
would decrease the resistance point (see Fig.~\ref{fig:trigger_size}),
but training-time attacks with large triggers are pointless.  The same
result can be achieved with an inference-time attack against the
\emph{unmodified} model, without any need for poisoning.  Adversarial
patches~\cite{bai2021inconspicuous,brown2017adversarial,pintor2023imagenet}
cover $2-10\%$ of the input and cause reliable misclassification.
Similarly, inference-time modifications using features from another class
don't require poisoning (see Section~\ref{sec:backdoors-vs-inference}).


\paragraphbe{Task injection.} 
It is easy to generate data for the primitive sub-task in many domains.
We use a random mask $M$ of size $s$ and pattern $P$ and create poisoned
inputs $x^\star=M {\cdot} x {+} (1{-}M) {\cdot} P$, $\forall x \in
\mathcal{X}$.  We further adapt them so as not to violate the constraints
on values or shapes and preserve the data type of $P$, as inputs may use
floats (e.g., images) or integers (e.g., tokenized texts)\textemdash see
details in Appendix~\ref{sec:wrapper}.  The pattern may still accidentally
contain features associated with a particular label.  We discuss how to
mitigate this effect in Appendix~\ref{sec:preventing_success}.

\paragraphbe{Backdoor interference.} 
The dataset used for measuring resistance to the primitive
sub-task may already be poisoned with another backdoor.  This does
not affect the measurements because models can learn multiple
backdoors~\cite{bagdasaryan2018backdoor}.  We observed that even
similarly-sized random triggers initialized with different seeds do not
interfere with each other.

%
%
%
%

\section{Discovering Resistant Configurations}

We now investigate how to select model configurations that have better
``natural'' resistance to backdoors without modifying the training
pipeline.

\subsection{Hyperparameters and Backdoors}

Hyperparameters such as the learning rate and batch size control
how the model learns both the main and backdoor tasks.  Recent
work~\cite{schwarzschild2021just} shows that efficacy of backdoor
attacks depends on the model's hyperparameters.  There is a tradeoff:
hyperparameters that render backdoor attacks ineffective can also have
a strong negative impact on the model's accuracy on its main task (Table
16 in~\cite{schwarzschild2021just}).

Searching for hyperparameters that result in a good model is a
standard part of configuring ML pipelines and can take place in the
experimentation zone~\cite{kreuzberger2022machine}.  Given a model
$\theta$, training data $\mathcal{D}$, training algorithm
$\mathcal{M}$, and a space of hyperparameters $\Lambda$, this search
solves an optimization problem: find a combination of hyperparameters
$\lambda \in \Lambda$ that optimizes a certain objective, e.g.,
maximize main task accuracy $\textbf{A}$ measured on the validation
dataset $\mathcal{D}_{val}$:
\begin{equation} \textbf{Main:}\;\;\;  \operatorname*{max}_{\lambda
\in \Lambda} \textbf{A}(\mathcal{D}_{val};\; \mathcal{M}(\mathcal{D},
\lambda)) \label{acc-objective} \end{equation}

Finally, selected hyperparameters $\lambda$ that optimize the
objective are evaluated on the unseen test dataset
$\mathcal{D}_{\text{test}}$ to provide an unbiased estimate of the
model's performance.

We want to balance \emph{two} objectives:
\textbf{(1)} the main objective, i.e., achieve high accuracy on the
main task, and \textbf{(2)} the resistance objective, i.e., increase
resistance of the model to poisoning-based backdoor attacks.

\subsection{Resistance Objective}
\label{sec:change_objective}

Maximizing the resistance objective should increase the
resistance point $p_{\star}^{\circ}$ of the model.  As described in
Section~\ref{sec:resistance_metric}, the model's backdoor accuracy
starts increasing when a $p_{\star}^{\circ}$ fraction of the dataset
is compromised.  For hyperparameter search, we convert this metric
into a learning objective: minimize the primitive sub-task accuracy
for $p_{\star} \geq p_{\star}^{\circ}$ measured on the fully poisoned
validation dataset $\mathcal{D}_{val}^{\star}$:
\begin{equation}
\textbf{Resistance:}\;\;\; \operatorname*{min}_{\lambda\in\Lambda}  \textbf{A}(\mathcal{D}_{val}^{\star};\;  \mathcal{M}(\mathcal{D}^{p_\star}, \lambda))
\label{resist-objective}
\end{equation}

Hyperparameters $\lambda_R$ that satisfy this objective and minimize
backdoor accuracy at $p_{\star}$ will push the resistance point
to $p_{\star,\lambda_R}^{\circ} \geq p_{\star}$.  To perform
the search, we only modify the training data, not the training
algorithm $\mathcal{M}$. We, first, randomly poison a $p_\star$
share of $\mathcal{D}$ with the primitive sub-task to obtain
$\mathcal{D}^{p_\star}$, and, second, create a new validation dataset
$\mathcal{D}_{val}^{\star}$ by fully poisoning $\mathcal{D}_{val}$ with
the primitive sub-task (see Appendix~\ref{sec:wrapper} for the details
of how we automate the data poisoning process).  We then measure the
new resistance point for the model with hyperparameters $\lambda_R$.

Actual backdoors are weaker (i.e., more poisoned training data are
required for the model to learn them), thus their resistance points
are higher.  The found hyperparameters increase the resistance
points for all backdoors, not just the primitive sub-task (see
Section~\ref{sec:exps_main}).

\subsection{Combining Accuracy and Resistance}
\label{sec:combining}

We modify hyperparameter search to jointly optimize for the main-task and
resistance objectives (see Figure~\ref{fig:overview}). Below, we explain
how to combine these objectives for different hyperparameter search tools.

\begin{figure}
    \centering
    \includegraphics[width=0.8\linewidth]{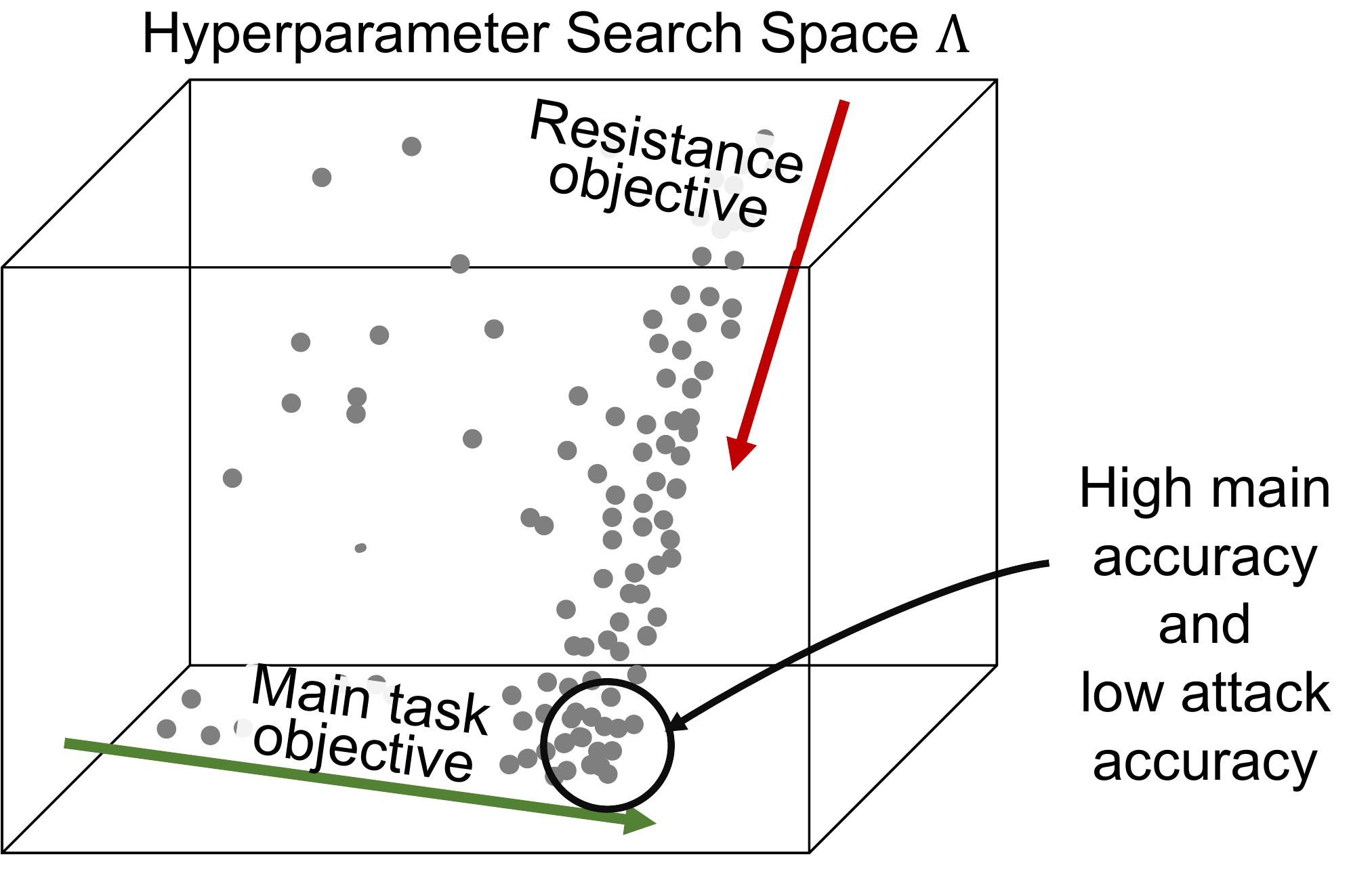}
    \caption{\textbf{Hyperparameter space with two objectives.}}
    \label{fig:hyperparam_space}
\end{figure}

\paragraphbe{Multiple objectives.}
A hyperparameter search tool capable of targeting multiple
objectives~\cite{parsa2020bayesian} can search for hyperparameters
that satisfy objectives~\ref{acc-objective} and~\ref{resist-objective}
together.  These objectives are based on the model's accuracy on different
validation datasets: respectively, $\mathcal{D}_{val}$ and poisoned
$\mathcal{D}_{val}^{\star}$ (see Figure~\ref{fig:hyperparam_space}).
Multi-objective optimization produces a Pareto frontier where one
objective can only be improved by harming the other~\cite{karl2022multi}.



\paragraphbe{Joint objective.} 
Some tools only allow one objective during the hyperparameter search,
e.g., ASHA~\cite{li2020system}. In this case, we can use a linear
combination of the two objectives balanced with coefficient $\alpha$:
\begin{equation}
\begin{split}
\label{eq:single}
\textbf{Joint:}\;\;\;\operatorname*{max}_{\lambda \in \Lambda} (&\alpha \cdot \textbf{A}(\mathcal{D}_{val};\;  \mathcal{M}(\mathcal{D}^{p_\star}, \lambda)) - \\
&(1-\alpha)\cdot \textbf{A}(\mathcal{D}_{val}^\star;\;  \mathcal{M}(\mathcal{D}^{p_\star}, \lambda)))
\end{split}
\end{equation}


Given an engineer 's specification of a permissible trade-off between 
the reduction
in the main-task accuracy $\vartriangle =
|\textbf{A}(\mathcal{D}_{val};\; \mathcal{M}(\mathcal{D}^{p_\star},
\lambda_1)) - \textbf{A}(\mathcal{D}_{val};\;
\mathcal{M}(\mathcal{D}^{p_\star}, \lambda_2))|$ and the drop
in the primitive sub-task accuracy $\vartriangle_\star =
|\textbf{A}(\mathcal{D}_{val}^\star;\;
\mathcal{M}(\mathcal{D}^{p_\star},
\lambda_1)){-}\textbf{A}(\mathcal{D}_{val}^\star;\;
\mathcal{M}(\mathcal{D}^{p_\star}, \lambda_2))|$ we can compute
$\alpha$ for the joint objective as:

\begin{equation}
\label{eq:alpha}
\alpha = \frac{\vartriangle_\star}{\vartriangle + \vartriangle_\star} 
\end{equation}



For example, if the engineer permits a drop in the main-task accuracy
of $\vartriangle = 3\%$ and requires a drop in the primitive sub-task
accuracy of $\vartriangle_\star = 100\%$, these metrics are balanced
by $\alpha = \frac{100}{3+100} \approx 0.967$.  Lower $\alpha$ allows
exploration of more backdoor-resistant configurations at a greater cost
to main-task accuracy; higher $\alpha$ increases main-task accuracy at
the cost of also increasing backdoor accuracy.


\paragraphbe{Joint validation dataset.} 
Some hyperparameter optimization tools do not allow computing accuracy
on different datasets $\mathcal{D}_{val}$ and $\mathcal{D}_{val}^\star$.
In this case, the engineer may assemble a single validation dataset
$\mathcal{D}_{val}^{\star}$ from the primitive sub-task inputs with their
original, correct labels.  High accuracy on $\mathcal{D}_{val}^{\star}$
indicates resistance to backdoor attacks and good main-task accuracy,
low accuracy indicates that the model is likely predicting backdoor labels
on inputs with the sub-task pattern (i.e., the backdoor is effective).
\begin{figure*}
\centering
\includegraphics[width=1.0\linewidth]{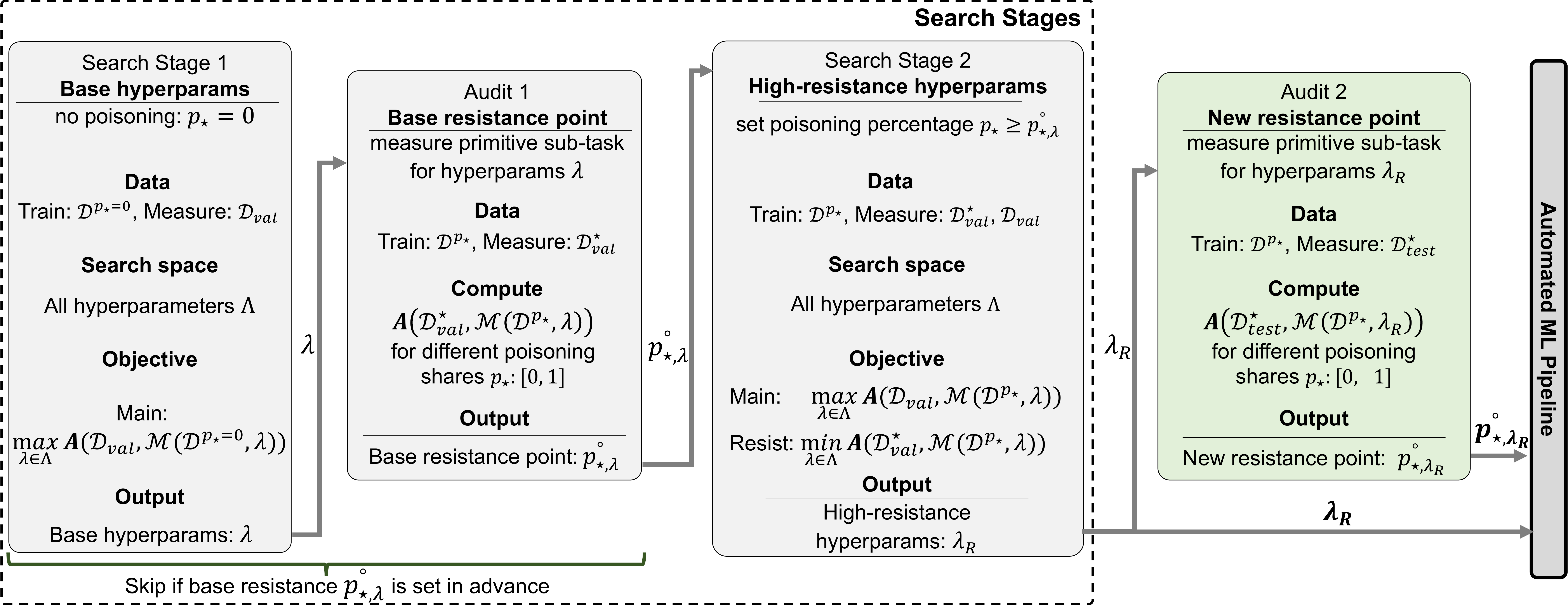}
\caption{\textbf{\method search for hyperparameters with higher 
backdoor resistance.}}
\label{fig:stages}
\end{figure*}

\subsection{\method Search}

We now develop a method to discover model configurations with high
natural backdoor resistance. First, we create a poisoned dataset
$\mathcal{D}^{p_\star}$ and set the fraction $p_\star$ of the data to
poison during hyperparameter search. The value of $p_\star$ may be based on
previous experiments or task-specific knowledge, but in general it
depends on the model's resistance point $p^\circ_\star$ which is not
known in advance. If $p_\star$ is too high, it is hard to balance the
objectives, i.e., preventing the model from learning the primitive
sub-task comes at a high cost to its main-task accuracy.  If $p_\star
{<} p^\circ_\star$, minimizing the resistance objective
$\textbf{A}(\mathcal{D}_{val}^{\star},
\mathcal{M}(\mathcal{D}^{p_\star}, \lambda))$ is trivial as the model
will not learn the backdoor anyway.

Therefore, we proceed in several stages, shown in Fig.~\ref{fig:stages}:
(1) search for the initial hyperparameters $\lambda$ to measure
the model's resistance to the primitive sub-task; (2) audit the base
resistance point $p^\circ_{\star,\lambda}$ for $\lambda$; (3) search
for high-resistance hyperparameters $\lambda_R$; and (4) audit
 the new resistance point $p^\circ_{\star,\lambda_R}$ and the
trade-off between $\lambda$ and $\lambda_R$ for the main task.  We use
validation sets $\mathcal{D}_{val}$ and $\mathcal{D}_{val}^\star$ in
the search stages and report model accuracy on $\mathcal{D}_{test}$
and $\mathcal{D}_{test}^\star$ in the final step.

\paragraphbe{Stage 1: Initial hyperparameter search.} 
We begin by finding base hyperparameters $\lambda$ that achieve good
main-task accuracy.  They may be known from previous runs,
otherwise we perform a generic hyperparameter search with the main objective
$\operatorname*{max}_{\lambda\in\Lambda} \textbf{A}(\mathcal{D}_{val};\;
\mathcal{M}(\mathcal{D}, \lambda))$.  


\paragraphbe{Audit 1: Base resistance point.} 
We poison the training dataset using different
percentages $p_\star \in [0, 1]$ and measure the primitive sub-task
accuracy: $A(\mathcal{D}_{val}^*;\; \mathcal{M}(\mathcal{D}^{p_\star},
\lambda))$ with hyperparameters $\lambda$ (see
Figure~\ref{fig:resistance_point}).
The resistance point $p^\circ_{\star, \lambda}$ corresponds to the
change in the second derivative or, for simplicity, the midpoint
between the maximum and minimum backdoor accuracy.
Fig.~\ref{fig:resistance_point} shows the curve for a CIFAR-10
model; the resistance point is $0.006\%$ ($3$ images).

\paragraphbe{Stage 2: Hyperaparameter search for high-resistance configurations.} 
Starting with the natural resistance point, we set $p_\star \geq
p^\circ_\star$ or $p_\star {=} k{\cdot} p^\circ_\star$ for some $k$
(we fix $k{=}2$, but searching for $k$ can be part of Stage 2).
We poison the training dataset $\mathcal{D}^{p_\star}$ and run either
a multi-objective search, or optimize a linear combination of main and
resistance objectives (Section~\ref{sec:combining}).  This produces
a set of hyperparameters and corresponding accuracies for the main
and primitive tasks that can be mapped as a Pareto frontier.  We then
either choose $\lambda_R$ manually, or set $\alpha$ and pick parameters
that maximize Equation~\ref{eq:single}.  Other hyperparameters are also
picked by varying the tradeoff between main-task and primitive sub-task
accuracy using $\alpha$ and Equation~\ref{eq:alpha}.  This stage outputs
hyperparameters $\lambda_R$.

\paragraphbe{Audit 2. New resistance point.} 
Using newly obtained $\lambda_R$, we compute the new resistance point
$p^\circ_{\star,\lambda_R}$ for the primitive sub-task on the
test set $\mathcal{D}_{test}^\star$.  We can also compute the resistance
point for other, realistic backdoors to ensure that they are higher
than $p^\circ_{\star,\lambda_R}$.

The resulting hyperparameters $\lambda_R$ can be used to configure the
ML pipeline.  The value of $p^\circ_{\star,\lambda_R}$ can be reused
for subsequent runs, skipping Stage 1.

\subsection{Limitations of Hyperparameter Search}

Exploring all $\lambda \in \Lambda$ is time-consuming and possibly
infinite because hyperparameters such as the learning rate can
be continuous.  Existing tools can navigate the search space using
complex analysis of model accuracy~\cite{akiba2019optuna} or early
stopping on less promising hyperparameters~\cite{li2020system}.
Furthermore, hyperparameter search can protect privacy of user
data~\cite{papernot2022hyperparameter}. Nevertheless, this is still an
\emph{empirical} method that can miss optimal hyperparameters.

\paragraphbe{Navigating the Pareto frontier.} 
Multi-objective hyperparameter search outputs a ``frontier,'' i.e., a set
of hyperparameter combinations mapped to two dimensions, main-task and
primitive sub-task accuracy.  If the frontier is vertical, the engineer
can maintain main-task accuracy while reducing accuracy on the primitive
sub-task (see Fig.~\ref{fig:frontier}); if diagonal, resistance comes at
a high cost in main-task accuracy (see Table~\ref{tab:target_resistance}).
The engineer can pick a point on the frontier depending on their specific
dataset and task.


\paragraphbe{Reducing computational overhead.} 
\method increases the time it takes to tune hyperparameters vs.\
``normal'' hyperparameter search, which only needs Stage 1.

There are several ways to reduce this overhead.  First, Stage 1 only needs
a baseline set and does not require many iterations.  Second, computing
reports 1 and 2 can use exponentially increasing poisoning percentages.
Finally, after the first run of multi-stage search, the discovered
resistance point $p^\circ_{\star,\lambda_R}$ can be re-used.  Tools such
as FFCV~\cite{leclerc2022ffcv}, Squirrel~\cite{2022squirrelcore}, and
Deep Lake~\cite{hambardzumyan2022deep}\textemdash even if not usable
in production due to complexity or incompatibility with the existing
data-loading pipelines\textemdash can help speed up data loading and
preprocessing during hyperparameter search.


\section{Practical Extensions}
\label{sec:practical_limitations}

We first explain how to incorporate regularization techniques into
resistance-boosting hyperparameter search, then discuss extensions to
federated learning and AutoML.

\subsection{Hyperparameter Selection}
\label{sec:params_booster}

Modern ML frameworks such as transformers~\cite{wolf2019huggingface}
support a large set of hyperparameters, giving the engineer many choices.
Therefore, the hyperparameter search space $\Lambda$ is specific to a
particular ML pipeline.  If we are thinking of poisoning-based backdoors
as learning from a sub-population~\cite{jagielski2020subpopulation,
wang2020attack}, to boost resistance we should focus on hyperparameters
that affect the learning of outliers.

\paragraphbe{Regularization as a defense.} 
Regularization helps models generalize and prevent
overfitting~\cite{tian2022comprehensive}.  Some regularization
methods are known to affect backdoor learning~\cite{cina2021backdoor,
demontis2018intriguing}, poisoning~\cite{carnerero2021regularization},
and unintended memorization~\cite{carlini2018secret} but we extend
this observation to all regularizations since even basic methods,
such as managing the learning rate and label perturbation, reduce
overfitting to small subsets of the training data and improve resistance
to backdoors.  A similar observation has been made in adversarial
training, where early stopping can be as effective as complex
defenses~\cite{rice2020overfitting}.

Many existing training-time backdoor defenses already rely on
regularization methods to prevent backdoor learning (even if not described
as such in the original papers):

\textit{Input perturbation}, e.g., filtering~\cite{chiang2020certified}
or perturbing~\cite{geiping2021doesn} training data helps the
model to not learn the backdoor trigger.  This is a form of data
augmentation~\cite{shorten2019survey}.

\textit{Gradient perturbation}, e.g., clipping or adding noise
to gradients~\cite{hong2020effectiveness}, ensures the model does
not get updated with the exact gradients.  This is very similar to
well-known generalization techniques~\cite{pascanu2013difficulty,
neelakantan2015adding}.

\textit{Label modification} helps the model to not learn the association
between backdoor features and labels~\cite{liu2022defending}, similar
to a standard regularization method~\cite{rolnick2017deep}.

\textit{Modification of the training mechanism} \textemdash a
broad range of defenses that prevent learning of backdoors\textemdash
is closely related to regularization.  For example, a state-of-the-art
defense RAB~\cite{weber2023rab} uses randomized smoothing, an existing
generalization method~\cite{duchi2012randomized} that provides robust
classification in the presence of noise~\cite{cohen2019certified}.

Therefore, an engineer who chooses hyperparameters to optimize in
their pipeline can leverage a rich toolbox of regularization techniques
(including basic ones like modifying the learning rate or early stopping)
to improve generalization and boost resistance to backdoors at the
same time.

\paragraphbe{Overfitting and outliers.} 
If we think of backdoored training data as outliers, we need to measure
the impact of resistance boosting not just on the average main-task
accuracy but also on outliers.  Furthermore, memorization of individual
inputs is important for long-tail performance~\cite{feldman2020does}.
Regularization techniques such as gradient shaping, while defending
against backdoors~\cite{hong2020effectiveness}, can negatively impact
underrepresented classes~\cite{bagdasaryan2019differential}.  We discuss
these issues in Section~\ref{sec:outliers}.

On the other hand, hyperparameters that enable longer training,
larger model size, or importance sampling~\cite{katharopoulos2018not}
significantly boost accuracy but cause memorization of training
data~\cite{tirumalamemorization}. Therefore, these techniques, along
with memorizing outliers, help the model learn backdoor tasks, negatively
affecting backdoor resistance.

\paragraphbe{Importance of individual hyperparameters.} 
Hyperparameter search can also provide insights on the specific
hyperparameters that help boost resistance for the exact task,
model, and data.  The search can also use analytics tools like
fANOVA~\cite{hutter2014efficient} to find hyperparameters that improve
main-task accuracy and backdoor resistance together.

Finally, if engineers are willing to add defenses to their pipelines,
our approach can facilitate finding the best hyperparameters for these
defenses.  For example, state-of-the-art defenses provide backdoor
resistance for up to $50\%$ of poisoned data~\cite{li2021anti} with
many new hyperparameters such as the loss threshold and isolation rate.
In this paper, we focus on natural resistance that can be achieved while
keeping the pipeline intact, and leave this extension to future work.

\subsection{AutoML and Neural Architecture Search} 
\label{sec:automl_extension}

Finding the best model architecture with neural architecture
search~\cite{elsken2019neural} or the most appropriate model
with AutoML~\cite{he2021automl, das2020amazon} is similar
to hyperparameter search.  For example, a popular AutoML tool
FLAML~\cite{Wang_FLAML_A_Fast_2021} is based on the hyperparameter search
tool Ray Tune~\cite{liaw2018tune}.  \method can be integrated into these
tools by adding a resistance objective.  We illustrate this with a toy
example in Section~\ref{sec:automl_results}.


\section{Evaluation}
\label{sec:exps_main}

We first measure models' resistance by computing the resistance
point of the primitive sub-task.  We then demonstrate how to search
for hyperparameters that prevent the model from learning that task
and measure how the resulting models resist actual backdoor attacks.
Finally, we evaluate extensions to federated learning and AutoML.

The data on which hyperparameter search is performed may not be clean,
i.e., it may already contain backdoored inputs.  This does not affect
\method's measurements of backdoor resistance, which are based on the
primitive sub-task.  When the triggers of the backdoor and the primitive
sub-task are different, we did not observe any interference.  If the
backdoor and sub-task triggers match (by accident or intentionally),
our audit method detects that accuracy of the primitive sub-task
is positive even when $p_\star{=}0$ (i.e., in the absence of
artificial poisoning) and regenerates the sub-task trigger (see
Appendix~\ref{sec:preventing_success}).

\renewcommand{\arraystretch}{1.5}
\begin{table}
    \centering
    \caption{\textbf{CIFAR-10 hyperparameter space.}}
    \label{tab:all_hyperparams}
    \setlength\tabcolsep{1.5pt}
    \vspace{0.2cm}
    \begin{tabular}{l@{\hskip 0.1in}lrr}
\multirow{2}{*}{Parameter} & \multirow{2}{*}{Available values} &
\multicolumn{2}{c}{Importance} \\
         \cmidrule(r){3-4}
        &  & Main & Primitive \\
    \midrule
    Batch size           & $[16, 32, 64, 128, 256]$ & 0.03 & 0.05   \\
    Decay                & log-interval $[10^{-7}, 10^{-3}]$  & 0.01 & 0.01\\
    Learning rate        & log-interval $[10^{-5}, 2]$   & 0.03 & 0.05\\
    Momentum             & interval $[0.1, 0.9]$    & 0.01 & 0.07\\
    Optimizer            & [SGD, Adam, Adadelta]    & 0.36 & 0.05\\
\multirow{2}{*}{Scheduler}            &  [StepLR, MultiStepLR,   & \multirow{2}{*}{0.01} & \multirow{2}{*}{0.01}  \\
                         & CosineAnnealingLR] \vspace{0.1cm}\\
                         \multicolumn{4}{c}{\textit{Trivial regularizations}\vspace{0.1cm}} \\
    Batch grad clip  & interval $[1, 10]$    & 0.01 & 0.04\\
    Batch grad noise & log-interval $[10^{-5}, 10^{-1}]$ & 0.02 & 0.03\\
    Label noise          &   interval $[0.0, 0.9]$  & 0.10 & 0.06\\
    \end{tabular}
\end{table}

\subsection{Experimental Setup}
\label{sec:exp_setup}
\label{sec:exp_backdoors_defs}
\label{sec:exp_datasets}

\paragraphbe{Hardware.} For hyperparameter search, we created a
distributed setup with three desktop GPU machines running Ubuntu
20.04. The first machine has two Nvidia Titan XP GPUs with 12GB memory
each, one RTX 6000 with 24GB memory and 64GB of RAM, the second machine
has 4 Nvidia GeForce RTX 2080 Ti with 12GB memory each and 128GB of RAM,
the third machine has 2 Nvidia 12GB GeForce RTX 2080 Ti and 256GB of
RAM. All machines are connected to a 1Gbps LAN network.

\paragraphbe{Software.}  
We configured the machines into a small Ray cluster~\cite{moritz2018ray}
to perform joint hyperparameter searches.  We use Ray Tune
v1.13~\cite{liaw2018tune} with Python 3.9 and PyTorch 1.12. Each model
trains on a dedicated GPU, 4 CPU processors, and no RAM restrictions.

We modified the \texttt{Backdoors101}
framework~\cite{bagdasaryan2020blind} implemented in
PyTorch~\cite{pytorch_link} and added new training tasks and a
dataset wrapper for injecting primitive backdoors into the data (see
Appendix~\ref{sec:wrapper}).  For experiments on the language task, we
used the HuggingFace Transformers~\cite{wolf2019huggingface} framework
v4.20.0 and added our wrapper.

\paragraphbe{Datasets.} 
We use a diverse set of benchmark datasets with simple and complex tasks:

\begin{itemize}

\item \textbf{FashionMNIST}~\cite{krizhevsky2009learning} -- 
this is a dataset of $28\times28$ images of various fashion items.
It contains $60,000$ training and $10,000$ test inputs.

\item \textbf{CELEBA-Smile}~\cite{liu2015faceattributes} -- 
this is a dataset of celebrity photos with 40 binary attributes each.
It contains $162,770$ training and $20,000$ test inputs.  We pick the
'smiling' attribute as our binary classification task.


\item \textbf{CIFAR-10}~\cite{krizhevsky2009learning} -- this is a
balanced dataset of diverse $32 \times 32$ images split into 10 classes,
with a total of $50,000$ training and $10,000$ test images.

\item \textbf{ImageNET LSVRC challenge}~\cite{ILSVRC15} -- we use the
full dataset that contains $1,281,167$ training and $100,000$ test $160
\times 160$ images labeled into $1,000$ classes.  The task is to predict
the correct label for each image.  We measure the Top-1 accuracy of
the prediction.

\item \textbf{RTE GLUE}~\cite{wang2018glue} -- this dataset supplies
two text fragments as input and provides a label indicating whether the
second fragment entails the first one.  The dataset contains $2,490$
training and $277$ test inputs.

\end{itemize}

\begin{figure}[tbp]
    \centering
    \includegraphics[width=1.0\linewidth]{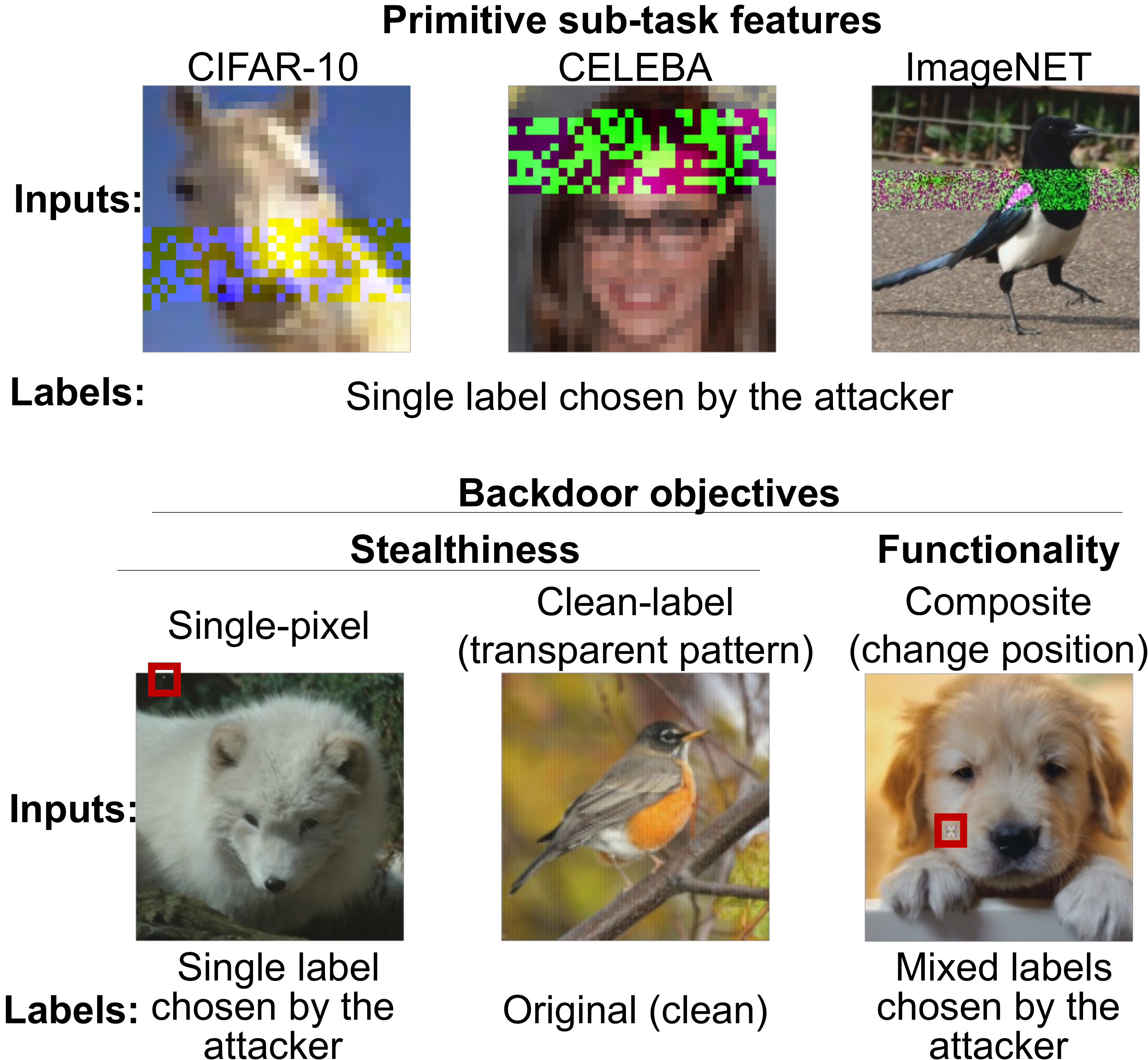}
\caption{\textbf{Primitive sub-tasks and backdoors.}}
    \label{fig:image_example}
\end{figure}

\begin{figure}[t]
    \centering
    \includegraphics[width=1.0\linewidth]{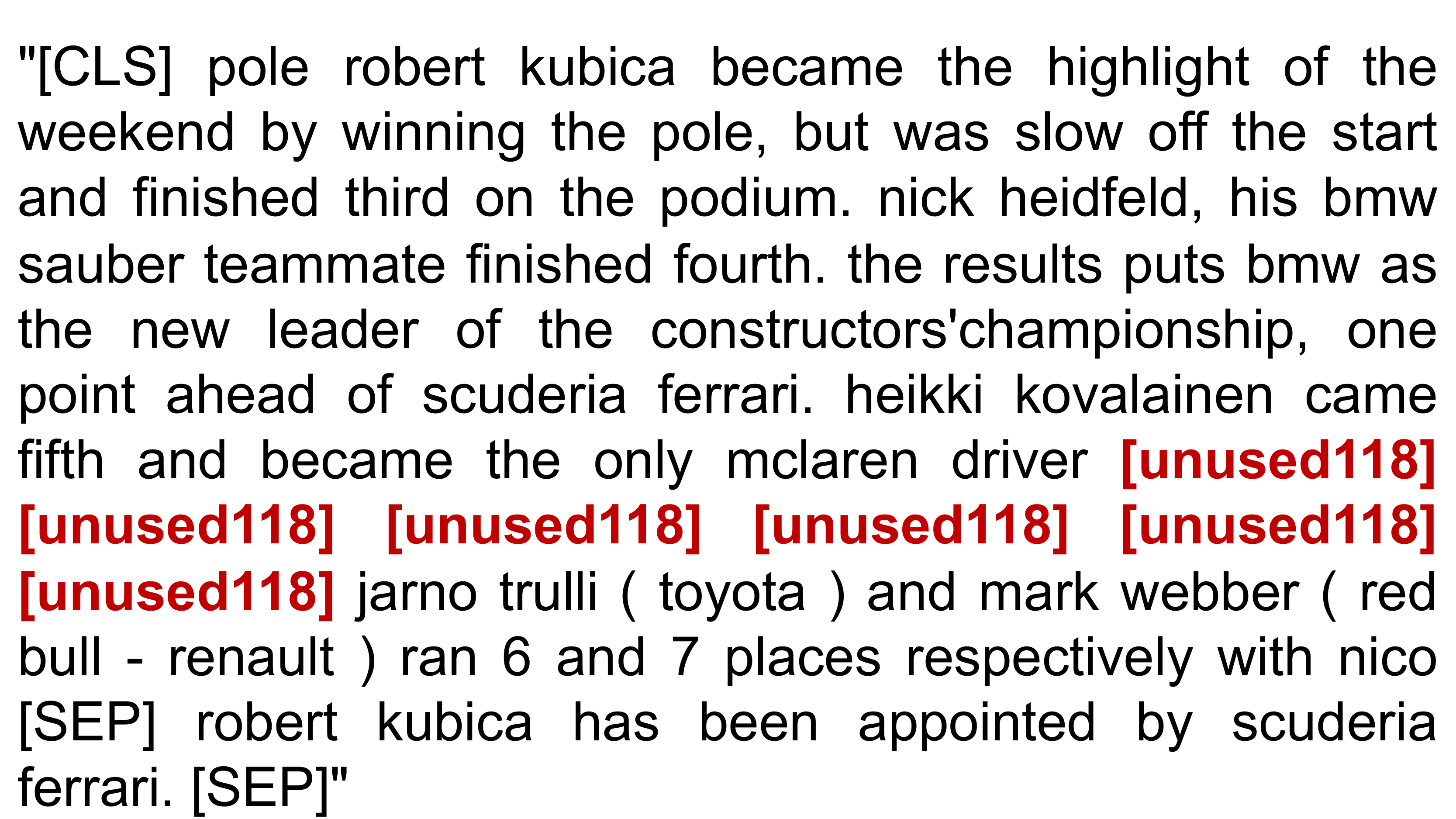}
\caption{\textbf{Primitive sub-task for text.}}
    \label{fig:text_example}
\end{figure}

For each dataset, we put $40\%$ of the test inputs into the
validation dataset $\mathcal{D}_{val}$, $60\%$ into the test dataset
$\mathcal{D}_{test}$.  The search stages use $\mathcal{D}_{val}$,
the final Report 2 is based on $\mathcal{D}_{test}$.  We recompute
resistance points with the non-resistant hyperparameters $\lambda$ on
the test set for the final-stage comparison with the high-resistance
hyperparameters $\lambda_R$.

\paragraphbe{Models.} 
For image and text classification, we use, respectively,
ResNet-18~\cite{he2016deep} with 11 million parameters and pre-trained
BERT~\cite{devlin2018bert} with 109 million parameters. For neural
architecture search, we assemble 2 convolutional layers followed by
two linear layers and parameterize layer configuration, dropout, and
activation functions.

To speed up training, we train CIFAR and CELEBA models for 10 epochs.
For CELEBA, we use a pre-trained model, CIFAR is trained from scratch.
To train ImageNet, we use FFCV~\cite{leclerc2022ffcv} to speed up dataset
loading and train the ResNet18 model from scratch for 16 epochs. Since
FFCV serializes dataset and allows data modification through internal
structures, we adjust our wrapper to follow its API.  Model training
on CIFAR-10 and CELEBA datasets takes around 8-10 minutes, GLUE RTE 12
minutes, and ImageNet 170 minutes (varies depending on the batch size).

\paragraphbe{Hyperparameter space.} 
We use standard hyperparameters such as the learning
rate and batch size, and add generic regularizations
such as label noise~\cite{rolnick2017deep}, batch-level
gradient clipping~\cite{pascanu2013difficulty} and gradient
noise~\cite{neelakantan2015adding}.  These are inexpensive,
easy-to-add functions supported by many frameworks, e.g.,
Transformers~\cite{wolf2019huggingface}.  Note that batch
gradient clipping is different from the expensive per-input
clipping used by DP-SGD~\cite{abadi2016deep} and gradient
shaping~\cite{hong2020effectiveness}.  Per-input clipping
slows down training and requires dedicated tools, e.g.,
Opacus~\cite{yousefpour2021opacus}.  See Table~\ref{tab:all_hyperparams}
for the full list of hyperparameters used in our image classification
experiments.  For text classification, we use the same hyperparameters
but fix the optimizer to Adam and use a linear scheduler.

\paragraphbe{Primitive sub-task.}
For each dataset, we generate a pattern for the primitive sub-task
using data augmentation (see Appendix~\ref{sec:wrapper}). We set the
coverage percentage $s{=}5\%$ because backdoor attacks with triggers
this large are pointless (see Sections~\ref{sec:backdoors-vs-inference}
and~\ref{sec:pointless}) and any real backdoor attack would use a smaller
trigger.


Examples of generated patterns are shown in
Figures~\ref{fig:image_example} and ~\ref{fig:text_example}.  Our
algorithm ignores input dimensionality and simply creates the mask as
a continuous sequence by treating the input as a 1D vector. Therefore,
the pattern on the image might cover only one of color channels, unlike
conventional pixel-pattern backdoors. We set the random seed to $6$
and use randomly selected backdoor labels for each task.

\paragraphbe{Backdoor objectives.} 
\label{sec-objectives}
As explained in Section~\ref{sec:primitive_task}, the purpose of the
primitive sub-task is to maximize the strength (i.e., minimize the
number of poisoned training inputs). To demonstrate that high-resistance
hyperparameters increase the resistance point for realistic backdoor
attacks, we evaluate models' resistance to stealthy and functional
backdoors.  As we show in Appendix~\ref{appx:backdoor_objectives}, attacks
that target these objectives require significantly more poisoned data
to be effective.  To provide a fair comparison, in our experiments we
use or create \emph{strengthened} versions of these attacks that require
less poisoned data, as follows.

\paragraphbe{Stealthiness + strength.} 
These backdoors aim to either hide the backdoor trigger, or preserve
label consistency:
\begin{itemize}
\item \textit{Imperceptible patterns}: this attack attempts to
modify image in the smallest possible way, measured by $l_0, l_1,
l_2, l_{\infty}$ distance or another metric~\cite{li2020backdoor}.
To increase the strength, we use a single-pixel attack similar to
RAB~\cite{weber2023rab} and place this pixel in the top right corner.

\item \textit{Clean label}: this attack also uses an
artificial, slightly transparent pattern but only attacks inputs
that have the backdoor label.  There are several clean-label
attacks~\cite{turner2019cleanlabel,souri2021sleeper}, but we use a very
strong Narcissus attack~\cite{zeng2022narcissus} that does not need a
pre-trained model or access to the data and increase $l_\infty$ norm
to $32/255$.
\end{itemize}

\paragraphbe{Functionality + strength.} 
These backdoors attempt to teach the model a more complex task by either
modifying the backdoor feature~\cite{wu2022just}, or adding logic on
backdoor labels:
\begin{itemize}
\item \textit{Composite pattern}: these attacks use complex
patterns that can be a physical object~\cite{Liu_2022_CVPR}
or focus on semantic features~\cite{bagdasaryan2018backdoor} and
transformations~\cite{wu2022just}. We strengthen the attack by dynamically
scaling and moving the pixel pattern across the image, simulating
the changing appearance of a physical object~\cite{composite2020,
Liu_2022_CVPR}.

\item \textit{Mixed labels}: this attack teaches the model
a complex task, to distinguish between different inputs
with the same trigger~\cite{bagdasaryan2020blind} or multiple
triggers~\cite{doan2022marksman}. To strengthen the attack, we split
inputs based on the true label and assign the backdoor label $0$ to the
first half (i.e., inputs whose true label is between $0$ and $L_{max}/2$
where $L_{max}$ is the max label value), $1$ to the rest.
\end{itemize}

\paragraphbe{Hyperparameter search settings.} 
For each task, we follow the \method method described in
Section~\ref{sec:params_booster}.  First, we perform hyperparameter
search with no poisoning (Search Stage 1) evaluating the model on
$\mathcal{D}_{val}$. Next, we find the base resistance point
$p_{\star,\lambda}^\circ$ for the primitive sub-task (Report 1) where
$50\%$ (midpoint) of the maximum sub-task accuracy is achieved.  We
then run hyperparameter search to find hyperparameters $\lambda_R$
that boost backdoor resistance (Search Stage 2) using $k{=}2$ for the
target poisoning $p_\star{=}k p_{\star,\lambda}^\circ$.  We finally
use the test set to $\mathcal{D}_{test}$ compute the resistance points
(Report 2) on base hyperparameters $\lambda$ and $\lambda_R$. For each
search stage, we train a multiple of $9$ models per the number of GPUs
available: $99$ models for Stage 1 and $360$ for Stage 2. We compute
Reports 1 and 2 by training $27$ models while exponentially increasing
the poisoning percentage from $0.001\%$ to $1\%$.

We use the Optuna~\cite{akiba2019optuna} optimization framework
integrated into the Ray Tune platform to perform multi-objective
search.  We additionally use an early stopping algorithm, Asynchronous
Successive Halving Algorithm (ASHA)~\cite{li2020system}, to stop
ImageNet training early.  Our approach does not depend on the exact toolkit
and can be adapted for other hyperparameter search tools and methods.
We did not notice significant changes when experimenting with other
optimization tools, e.g. HyperOpt~\cite{bergstra2013hyperopt} and
SigOpt~\cite{dewancker2016bayesian}.

\begin{figure*}
    \centering
    \includegraphics[width=0.7\linewidth]{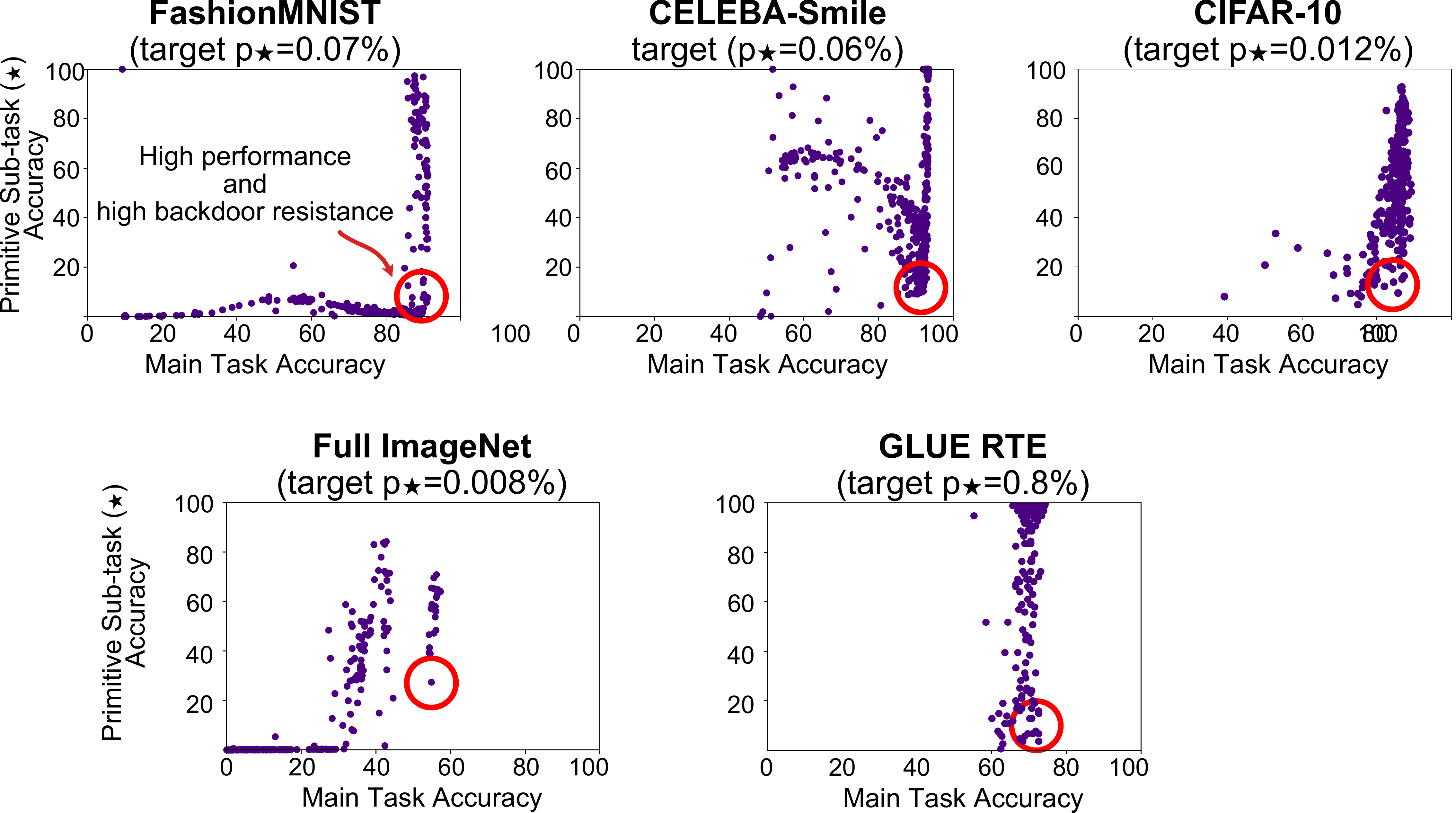}
\caption{\textbf{Hyperparameter Pareto frontier.}}
    \label{fig:frontier}
\end{figure*}

\subsection{Resistance to Primitive Sub-task}

\paragraphbe{Natural resistance.} 
Table~\ref{tab:main_results} shows main-task accuracy for different tasks
for base configurations and the corresponding resistance points. Simpler
tasks like FashionMNIST (which uses a small network) and CELEBA (which
only performs binary classification with a pre-trained ResNet model)
have higher resistance points ($0.27\%$ and $0.031\%$), i.e., require
a larger fraction of the training data to be poisoned to learn even the
primitive sub-task.  More complex tasks CIFAR-10 and ImageNET are learned
from scratch and appear easier to poison ($p=0.006\%$ and $0.004\%$).
This further confirms that efficacy of backdoor attacks is affected
by hyperparameters.

\paragraphbe{High-resistance configurations.} 
Using $p_\star{=}2 p_{\star,\lambda}^\circ$, \method runs Search
Stage 2 to find hyperparameters that balance the main and resistance
objectives. Figure~\ref{fig:frontier} shows that, given a fixed poisoning
percentage $p$, accuracy of the primitive sub-task on validation set
$\mathcal{D}_{val}$ varies significantly depending on the hyperparameters.
high-resistance hyperparameters reduce the sub-task accuracy to almost
$0\%$ for all tasks, except ImageNet where it drops from $70\%$ to $25\%$.

We then evaluate the newly found hyperparameters on
$\mathcal{D}_{test}$. Table~\ref{tab:main_results} shows the impact on
main-task accuracy and the resistance points.  For the more complex tasks,
high-resistance hyperparameters reduce main-task accuracy more but
also boost backdoor resistance by a factor of 2.5-5$\times$.

\renewcommand{\arraystretch}{1.5}
\begin{table}[tbp]
    \centering
\caption{\textbf{Higher backdoor resistance has moderate impact on
model accuracy.}}
    \vspace{0.2cm}
    \label{tab:main_results}
    \setlength\tabcolsep{2.0pt}
    \begin{tabular}{l@{\hskip 0.1in}c|c}
\multirow{3}{*}{Dataset} & Main task & Resistance point \\
& \multicolumn{1}{c|}{accuracy} &\% of dataset $\mathcal{D}$ \\
& $\textbf{A}_{\lambda} \rightarrow \textbf{A}_{\lambda_R}$ 
& $p_{\star,\lambda}^{\circ} \rightarrow p_{\star,\lambda_R}^{\circ}$ \\
     \midrule
FashionMNIST & $92.7 {\rightarrow}     90.5(\text{-}2.2)$             
& $0.035 {\rightarrow}0.120(\times \textbf{3.4})$ \\     
CELEBA-Smile & $92.8 {\rightarrow}     91.4(\text{-}1.4)$             
& $0.031 {\rightarrow} 0.121(\times \textbf{3.9})$ \\ 
CIFAR-10 & $89.3 {\rightarrow}     86.5(\text{-}2.8)$             
& $0.006 {\rightarrow}0.032(\times \textbf{5.3})$ \\     
ImageNet & $57.9 {\rightarrow}     54.6(\text{-}3.3)$           
& $0.004 {\rightarrow} 0.010(\times\textbf{2.5})$  
\\
GLUE RTE
& $70.2 {\rightarrow} 68.9(\text{-}1.3)$  
& $0.402 {\rightarrow}1.566(\times \textbf{3.9})$ \\

    \end{tabular}
\end{table}

\paragraphbe{Hyperparameter importance.} 
Table~\ref{tab:all_hyperparams} shows the importance of different
hyperparameters for backdoor resistance of CIFAR-10, computed using
fANOVA~\cite{hutter2014efficient} over Search Stage 2 results.
Resistance is sensitive to the batch size, learning rate, momentum,
optimizer, and label noise.

\subsection{Resistance to Actual Backdoors}

Next, we measure if higher resistance to the primitive sub-task also
increases resistance to stealthy and complex backdoors.  As we show
Attacks that prioritize stealthiness or functionality already require
higher poisoning percentages (Appendix~\ref{appx:backdoor_objectives}).
We modify the attacks to make them stronger (i.e., require less
poisoning) as described in Section~\ref{sec-objectives} while keeping
their stealthiness and/or functionality objectives.  We use CIFAR-10
for these experiments.

Table~\ref{tab:objectives} shows that the required poisoning percentage
is higher for the stealthy and complex backdoors.  high-resistance
hyperparameters increase these resistance points even further.  Since
their initial values were higher, the relative increase is smaller than
for the primitive sub-task.


If the engineer has a particular resistance metric in mind (i.e.,
maximum fraction of the training data that may be compromised),
\method can help achieve it\textemdash at a higher cost to main-task
accuracy\textemdash by executing Search Stage 2 with the target
resistance point.  Table~\ref{tab:target_resistance} shows the results.
For example, $10\times$ boost comes at the cost of a $15\%$ reduction in
main-task accuracy; $100\times$ costs $40\%$ accuracy.  These costs are
comparable to certified backdoor robustness~\cite{chiang2020certified,
wang2020certifying, weber2023rab} which only certifies against backdoors
of 1-4 pixels.

\paragraphbe{Adaptive attacks.} 
To craft an adaptive attack requiring even fewer inputs than
our primitive sub-task, the adversary may increase the trigger
size (see Figure~\ref{fig:trigger_size}).  As explained in
Sections~\ref{sec:backdoors-vs-inference} and~\ref{sec:pointless},
this attack would be pointless.  With triggers this large,
poisoning is unnecessary since the same effect can be
achieved with a pure inference-time attack based on adversarial
patches~\cite{pintor2023imagenet,bai2021inconspicuous} or injection
of features from another class.  Attacks that don't support arbitrary
triggers because they rely on the advance knowledge of the model
(e.g. derive triggers from adversarial examples or learned features) or
natural backdoors already present in benign data are a different threat,
outside our scope.


Another example of an attack that requires few inputs is a recent
few-shot backdoor attack~\cite{hayase2022few}.  It assumes the adversary
has access to at least $25\%$ of the training dataset and only works on
certain model architectures, but is yet order-of-magnitude weaker than
our primitive sub-task (e.g., for CIFAR-10 it requires $0.08\%$ of the
two-label dataset vs. $0.006\%$ of the full dataset for our sub-task).

\renewcommand{\arraystretch}{1.5}
\begin{table}[t!]
    \centering
\caption{\textbf{Higher resistance to the primitive sub-task increases
resistance to different backdoors (CIFAR-10).}}
    \label{tab:objectives}
    \vspace{0.2cm}
    \setlength\tabcolsep{1.5pt}
    \begin{tabular}{lc|c}
\multirow{3}{*}{Task} & Main task & Resistance point \\
& \multicolumn{1}{c|}{accuracy} &\% of dataset $\mathcal{D}$ \\
& $\textbf{A}_{\lambda} \rightarrow \textbf{A}_{\lambda_R}$ 
& $p_{\lambda}^{\circ} \rightarrow p_{\lambda_R}^{\circ}$ \\
        \midrule
\multicolumn{3}{l}{\textit{Strength only}} \vspace{0.05cm}\\
\;        Primitive
& $89.3 {\rightarrow}     86.5(\text{-}2.8)$             
& $0.006 {\rightarrow}0.032(\times \textbf{5.3})$ \vspace{0.1cm} \\ 
\multicolumn{3}{l}{\textit{Stealthiness + Strength}} \vspace{0.05cm} \\
\; Single dot~\cite{weber2023rab} &  $89.6 {\rightarrow} 87.1    (\text{-}2.5)$ 
& $0.194 {\rightarrow} 0.588   (\times\textbf{3.0})$  \\
\; Clean label~\cite{zeng2022narcissus} &  $89.8 {\rightarrow} 86.3    (\text{-}3.5)$ 
& $0.018 {\rightarrow} 0.066   (\times\textbf{3.7})$ \vspace{0.1cm}\\

\multicolumn{3}{l}{\textit{Functionality + Strength}} 
\vspace{0.05cm} \\
\; Composite~\cite{composite2020} & $89.2 {\rightarrow} 86.1 (\text{-}3.1)$  
& $ 0.256 {\rightarrow} 0.776(\times \textbf{3.0})$ 
\\
\; Mixed labels~\cite{bagdasaryan2020blind} & $89.3 {\rightarrow} 86.0 (\text{-}3.3)$  
& $ 0.053 {\rightarrow} 0.143 (\times \textbf{2.7})$ 

\end{tabular}
\end{table}

\renewcommand{\arraystretch}{1.5}
\begin{table}[tbp]
    \centering
    \caption{\textbf{Higher backdoor resistance impacts accuracy (CIFAR-10).}}
    \label{tab:target_resistance}
    \vspace{0.2cm}
    \begin{tabular}{cc|c}
Main task & \multicolumn{2}{c}{Resistance point,
\% of dataset $\mathcal{D}$}\\
\cmidrule{2-3}
 \multicolumn{1}{c}{accuracy $\textbf{A}$}
&Primitive & 1-pixel\\
     \midrule
      \multicolumn{1}{l}{$89.3$}             
     & \multicolumn{1}{l|}{\,$0.006$} 
     & \multicolumn{1}{l}{\:\,$0.19$}
     \\     
 $87.2\;\:(\text{-}2.1)$             
& $0.015\;\:\;\:\;(\times \textbf{2.5})$ 
& $0.58\;\:(\times \textbf{3.0})$ \\ 


$ 74.5(\text{-}14.8)$             
& $0.056\;\:\;\:\;(\times \textbf{9.3})$ 
& $1.78\;\:(\times \textbf{9.4})$\\     

$46.9(\text{-}42.4)$            
& \,$0.624(\times \textbf{104.0})$     
& $9.41({\times} \textbf{49.5})$ \\
\end{tabular}
\end{table}

\subsection{Extensions}
\label{sec:fedlearn_exps}
\label{sec:automl_results}
\label{sec:fl_extension}

\paragraphbe{Federated learning.} Federated learning trains models on
users' devices and collects only their weights to create a joint
global model.  Resistance of federated models to poisoning-based
backdoor attacks depends on training
hyperparameters~\cite{shejwalkar2022back}.

We use standard federated averaging~\cite{fedlearn_1}.  At every round
$q{=}[1\mathrel{{.}\,{.}}\nobreak G]$, we distribute the current global
model $\theta^g_q$ to a random user set $U$. Each user $i \in U$ trains
a local model $\theta^i_q$ for $l$ local epochs and computes an update
$\theta^g_q$.  The new global model is $\theta^g_{q{+}1} {=} \theta^g_q
{+} \eta \sum^{U}_{i{=}1}{(\theta^g_q{-}\theta_{q}^i)}$, where $\eta$
is the global learning rate.

Poisoning takes place on client devices, but we assume
that the attacker can only control the data and not the model
training~\cite{shejwalkar2022back}.  As the inflection point, we use the
fraction of compromised users needed to make the backdoor effective in
the global model $\theta^G$ at the end of training.
 
We use the FashionMNIST
and CIFAR-10 datasets split into $500$ and $100$ users with an equal
number of images per user (non-iid case).  For hyperparameter search,
we use the hyperparameters from Table~\ref{tab:all_hyperparams}
and add round size $M= [5\mathrel{{.}\,{.}}\nobreak 20]$, global learning rate
$\eta{=} (10^{{-}5}\mathrel{{.}\,{.}}\nobreak 10)$, and the number
of local epochs $l {=} [1\mathrel{{.}\,{.}}\nobreak 5]$.  We further add server-level weight clipping and
noise vectors~\cite{fedlearn_dp} (they don't impact the local training
pipeline).  Table~\ref{tab:fl_result} shows that \method significantly
boosts backdoor resistance.

\renewcommand{\arraystretch}{1.5}
\begin{table}[tbp]
    \centering
\caption{\textbf{High backdoor resistance in federated learning.}}
    \label{tab:fl_result}
    \vspace{0.2cm}
    \setlength\tabcolsep{1.5pt}
    \begin{tabular}{l@{\hskip 0.15in}c|c}
\multirow{2}{*}{Dataset}& Main task & Resistance point \\
& accuracy & \% of participants \\

& $\textbf{A}_{\lambda} \rightarrow \textbf{A}_{\lambda_R}$ & $p_{\star,\lambda}^{\circ}
\rightarrow p_{\star,\lambda_R}^{\circ}$ \\
     \midrule
FashionMNIST 
& $83.5 {\rightarrow} 80.4(\text{-}3.1)$
& $ 3.6 {\rightarrow}\;\; 21\; (\times\textbf{5.8})$
\\
CIFAR-10 
& $69.8 {\rightarrow} 67.4(\text{-}2.4)$
& $ 4.0 {\rightarrow}\;\;\;\: 9\; (\times\textbf{2.2})$
\\
\end{tabular}
\end{table}

\paragraphbe{Neural architecture search and AutoML.} 
We illustrate how \method works with neural architecture search on a small
example.  For FashionMNIST, we search for an architecture that satisfies
both the main-accuracy and resistance objectives, changing the model's
activation function, dropout, and convolution-layer hyperparameters,
e.g., stride, kernel, and linear layer.  Table~\ref{tab:automl} shows
the results.

\renewcommand{\arraystretch}{1.5}
\begin{table}[tbp]
    \centering
    \caption{\textbf{High backdoor resistance with AutoML (FashionMNIST).}}
    \label{tab:automl}
    \vspace{0.2cm}
    \setlength\tabcolsep{1.5pt}
    \begin{tabular}{lc|c}
\multirow{3}{*}{Search Space} & Main task & Resistance point \\
& accuracy &\% of dataset $\mathcal{D}$ \\
& $\textbf{A}_{\lambda} \rightarrow \textbf{A}_{\lambda_R}$ 
& $p_{\star,\lambda}^{\circ} \rightarrow p_{\star,\lambda_R}^{\circ}$ \\
        \midrule
Hyperparams $\Lambda$
& $92.7 {\rightarrow}     90.5(\text{-}2.2)$       
& $0.035 {\rightarrow}0.120(\times \textbf{3.4})$ \\     
AutoML
&  \multirow{1}{*}{\;\:$92.0 {\rightarrow} 91.1    (\text{-}0.9)$ }
& \multirow{1}{*}{\;\:$0.028 {\rightarrow} 0.192   (\times\textbf{6.9})$ }
\\
\end{tabular}
\end{table}

\subsection{Impact on the Long Tail}
\label{sec:outliers}

Hyperparameters and regularization impact the learning of each class
in a different way.  DP-SGD (a mix of per-input gradient clipping and
Gaussian noise)~\cite{abadi2016deep} is known to have disparate impact
on underrepresented subgroups~\cite{bagdasaryan2019differential}. We
follow the intuition derived in the edge-case~\cite{wang2020attack}
and subpopulation~\cite{jagielski2020subpopulation} attacks\textemdash
a strong backdoor targets tail inputs.  Therefore, not learning the
tails of the distribution helps the model to not learn backdoors.
This suggests that making the model more resistant to backdoors will
decrease its accuracy on underrepresented classes.

\begin{figure}[t]
    \centering
    \includegraphics[width=1.0\linewidth]{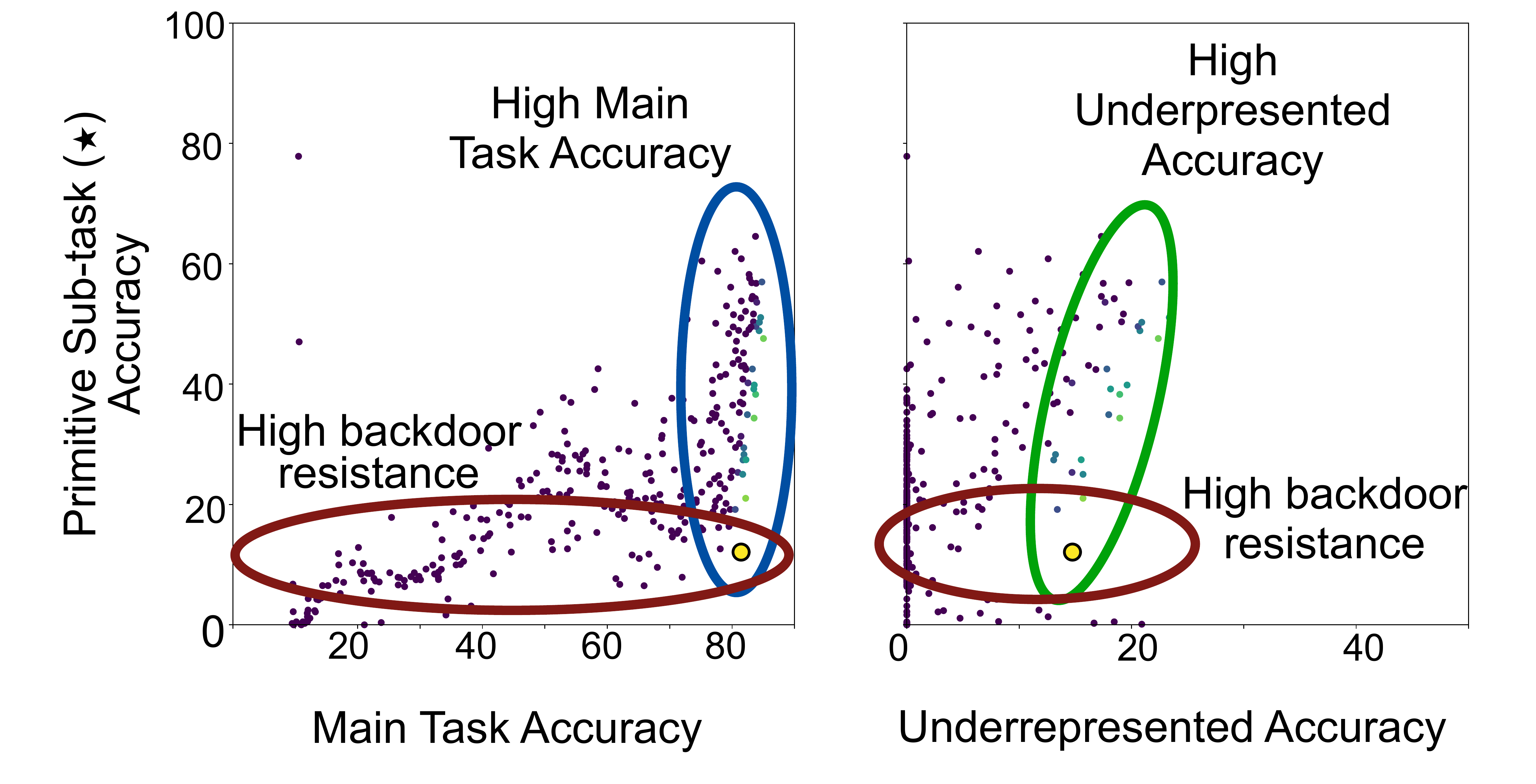}
\caption{\textbf{Impact on accuracy for underrepresented subgroups
(CIFAR-10 with downsampled class "cat").}}
    \label{fig:fairness}
\end{figure}

We use the CIFAR-10 dataset, downsample class $cat$ by $90\%$ to only
$500$ images, and perform hyperparameter search for backdoor resistance.
Figure~\ref{fig:fairness} demonstrates that high-resistance
hyperparameters decrease accuracy on the underrepresented class.
This problem can also be addressed in mechanism-agnostic way by
collecting a more balanced dataset or adding a separate fairness
objective~\cite{karl2022multi}.

\section{Conclusions and Future Work}

Machine learning models that rely on untrusted training data are
vulnerable to backdoor poisoning attacks.  Deployment of defenses from the
research literature requires expert knowledge and extensive modifications
to training pipelines.  Fortunately, efficacy of backdoor attacks can
be reduced by training with appropriate hyperparameters (in particular,
regularizations) while keeping the training pipeline intact.

In this paper, we took the perspective of a pragmatic ML engineer who
wants to (1) audit their existing pipeline to estimate its resistance
to backdoor poisoning, and (2) discover training configurations with
higher resistance.  For auditing, we proposed a metric
to estimate models' resistance to unknown backdoor attacks.  We then
developed and evaluated \method, a new multi-objective variant of
hyperparameter search\textemdash already a standard tool in practical
ML deployments\textemdash to help find hyperparameters that increase the
model's resistance to backdoor poisoning while maintaining its accuracy
on the main task.

\method helps find the balance between security and accuracy without
disruptive pipeline modifications.  It can also inform new backdoor
defenses that leverage existing regularization methods.  Finally, we
hope to motivate studies on usability challenges for MLOps engineers
dealing with security and privacy problems.

\section*{Acknowledgments}

This research was supported in part by the NSF grant 1916717, a Google
Faculty Research Award, and an Apple Scholars in AI/ML PhD fellowship
to Bagdasaryan.

\small
\bibliographystyle{IEEEtranS}
\bibliography{main}

\appendices
\section{Backdoor Objectives}
\label{appx:backdoor_objectives}

Our primitive sub-task aims to act as a lower bound on the fraction of the
training dataset that must be poisoned for any backdoor to be effective.
In addition to the poisoning percentage, backdoor attacks may have other
objectives.  Attacks may try to achieve trigger stealthiness, label
consistency, complexity, or defense evasion.  Any of these objectives
is an additional constraint that decreases the signal to be learned,
thus the attack should require a higher poisoning percentage than the
primitive sub-task (which is visually noticeable, label-inconsistent,
and easily detectable). Table~\ref{tab:backdoorreported} summarizes
the reported poisoning percentages for different attacks.

\subsection{Stealthy backdoors}

\paragraphbe{Small or imperceptible triggers.} 
Attacks like BadNets~\cite{badnets} use a pixel pattern
to change the label. The model must learn to prefer this
pattern over ``normal'' features to classify backdoored
inputs to an attacker-chosen label.  The pattern can be very
small, e.g., a single pixel~\cite{bagdasaryan2020blind},
or imperceptible~\cite{chen2017targeted}.  The pattern for our
primitive sub-task is larger and fully covers more normal features.
In Section~\ref{sec:exps_main}, we show that smaller patterns indeed
make learning the backdoor harder.

\paragraphbe{Clean labels.} 
Clean-label backdoor attacks~\cite{geiping2021witches,
wallace2020customizing, turner2019cleanlabel, liu2020reflection}
create a backdoored tuple $(x^*_c, y^*)$ that looks plausible, i.e.,
a human would assign label $y^*$ to input $x^*_c$.  An attacker uses a
clean input $x_c$ that has label $y^*$ and blends $x^*$, i.e., input $x$
with the trigger applied, into $x_c$ to obtain $x_c^*$ that looks like
$x_c$. Some attacks work well on pretrained models~\cite{cina2021backdoor,
geiping2021witches} when a part of the model, e.g., the embedding
layer $\theta_{emb}$, is frozen.  These attacks use techniques like
gradient alignment~\cite{souri2021sleeper} or projected gradient
descent~\cite{saha2019hidden} to generate $x^*_c$ that produces the
same embedding $\theta_{emb}(x^*_c) {=} \theta_{emb}(x^*)$ as the
backdoored image $x^*$ while it is similar to $x_c$ within some distance
$\varepsilon$, e.g., $||x^*_c {-} x_c||_{2} {=} \varepsilon$.  In other
words, when applied to input $x^*_c$, the model ``sees'' a backdoored
input $x^*$ with the trigger while a human sees a clean input $x_c$.



Stealthiness of the attack, i.e., the value of $\varepsilon$, is
not a useful metric in our setting because data inspection is a
(disruptive) defense and requires integration into the ML pipeline.
Attacks that overlay the entire image with some perturbation, e.g.,
reflection~\cite{liu2020reflection} or a ramp signal~\cite{barni2019new},
are also limited by the amount of perturbation.  We can maximize
$\varepsilon$ to make the attack strong with $\varepsilon{=}x^*{-}x_c$
that converts the blended input $x^*_c$ back to $x^*$, making the attack
label-inconsistent.  Therefore, clean-label attacks are weaker (i.e.,
require a higher poisoning percentage) because they need to satisfy a
tight $\varepsilon$ budget to remain stealthy.  Indeed, values reported
by the papers~\cite{souri2021sleeper, turner2019cleanlabel} show that
$0.1{-}1\%$ of the training data must be poisoned in order for the
attack to become effective.  By contrast, Section~\ref{sec:exps_main}
shows that the primitive sub-task is learned with as little as $0.005\%$
of the dataset compromised.

\subsection{Functional backdoors}
 
Some attacks use composite~\cite{composite2020},
physical~\cite{bagdasaryan2020blind} or semantic~\cite{wu2022just}
features as triggers. These attacks also require a larger fraction
of the training data to be poisoned (e.g., $10\%$) in order to
become effective~\cite{composite2020}. Complex attacks that inject
an extra task (e.g., a meta-task~\cite{bagdasaryan2022spinning})
are only effective when $50\%$ of the training data is compromised.
In Section~\ref{sec:exps_main}, we show that even basic functionalities
like dynamic movement of the pattern~\cite{salem2022dynamic} and label
shift~\cite{doan2022marksman} are harder for the model to learn than
the primitive sub-task.

\subsection{Targeted and sample-specific attacks} 
\label{sec:targeted_attacks}

Some attacks do not teach the model a \emph{generalizable}
backdoor task but instead cause misclassification for a specific
input~\cite{geiping2021witches, shafahi2018poison}.  These attacks
focus on memorization of a given input and may require only a single
compromised example~\cite{koh2017understanding}\textemdash but this
learning does not generalize.  

On the other hand, sample-specific attacks~\cite{li2021invisible,
zhang2022poison} generate individual triggers for each input
using steganography~\cite{zhang2021brief}. These attacks are more
complex (because the model needs to learn the task corresponding to
sample-specific trigger generation) and require a larger fraction of
the training data to be poisoned.




\renewcommand{\arraystretch}{1.0}
\begin{table}[tbp]
    \centering
    \caption{\textbf{Reported poisoning percentages}}
    \label{tab:backdoorreported}
    \vspace{1ex}
    \begin{tabular}{l@{\hskip 0.15in}rr}
Backdoor & $p_b$ & Domain \\
     \midrule

\textit{Stealthy}
\\
\;\; BadNets~\cite{badnets} & 10.00\% & Images  \\
\;\; BadNL~\cite{chen2020badnl} & 3.00\% & Text \\
\;\; Clean image~\cite{chen2023cleanimage} & 1.50\% & Images \\
\;\; Narcissus~\cite{zeng2022narcissus} & 0.05\% & Images \\
\;\; Poison-ink~\cite{zhang2022poison} & 3.00\% & Images \\
\;\; Sentiment~\cite{dai2019backdoor} & 0.50\% & Text \\
\;\; Sleeper~\cite{souri2021sleeper} & 0.10\% & Images\\
\;\; Sample-specific~\cite{li2021invisible} & 10.00\% & Images \\
\;\; Refool~\cite{liu2020reflection} & 1.00\% & Images \vspace{0.2cm}\\

\textit{Functional} \\
\;\; Composite~\cite{composite2020} & 8.30\% & Images\\
\;\; Dynamic~\cite{salem2022dynamic} & 10.00\% & Images\\
\;\; LLM Spinning~\cite{bagdasaryan2022spinning} & 50.00\% & Text \\
\;\; Marksman~\cite{doan2022marksman} & 5.00\% & Images\\
\;\; Rotation~\cite{wu2022just} & 1.00\% & Images \\
\;\; Physical~\cite{bagdasaryan2020blind} & 33.00\% & Images\\

\end{tabular}
\end{table}

\section{Automation for Dataset Poisoning}
\label{sec:wrapper}

Even when using a third-party service with fixed API to train their
models, engineers control their data, in particular the training and
validation datasets.  Finding a resistance point requires a training
dataset $\mathcal{D}^{p_\star}$ poisoned with the primitive sub-task
at a specific percentage $p_\star$.  How to carry out this poisoning
depends on the framework but can also be done when the dataset is created.

We use a wrapper around the dataset that automatically injects
poisoned data at a specified percentage: $\mathcal{D}^{p_\star} {=}
\texttt{Attack}(\mathcal{D}, p_\star)$.  An engineer can either integrate
this wrapper, or generate a new dataset.  The poisoned dataset is used
during the hyperparameter search.  The production pipeline will use the
original, unmodified dataset.

\begin{algorithm}[t!]
    \linespread{0.9}\selectfont

    \caption{Dataset wrapper for primitive sub-task.}
    \small
        \label{alg:poison}
        \vspace{0.2cm}
    \textbf{INPUTS:} dataset $\mathcal{D}$, attack percentage
    $p_\star$, patch size $s$, backdoor label $y^*$.\\
    \vspace{0.1cm}
    \SetKwProg{Class}{class}{}{end class}
    \SetKwProg{Def}{def}{}{end def}
    
    \Class{AttackDataset}{
    \textbf{fields}: $\mathcal{D}, s, p_\star, y^*, \text{ indices }
    I^*, \text{ mask }M, \text{ pattern } P$\\
    \Def{\_\_init\_\_($\mathcal{D}$, $s$, $p_\star$, $y^*$)}{
        \vspace{0.1cm}
        \textit{\# compute poisoned indices} \\ 
        $I^* \leftarrow \texttt{sample}(\mathcal{D}, p_\star)$ \\
        $M, P \leftarrow \texttt{create\_patch}(\mathcal{D}, s)$ \\
    }
    \vspace{0.1cm}
    \Def{\_\_get\_\_(index $i$)}{
            $(x, y) \leftarrow \mathcal{D}[i]$\\
            \uIf{$i \in I^*$}{
                \KwRet \texttt{apply\_pattern}($x$)\\
            }
            \Else{
                \KwRet $(x, y)$\\
            }
    }
    \vspace{0.1cm}
    \Def{create\_patch($\mathcal{D}$, $s$)}{
        \textit{\# assume inputs are 1D, get stats} 
        
        $x_{max}, x_{min}, x_{len} \leftarrow \mathcal{D}$

        $m_{start} \leftarrow  \texttt{rand\_int}(0, l - s*x_{len})$ \\
        \vspace{0.1cm}
        \textit{\# Make a mask}
        
        $ M(i) = \begin{cases} 1, & \text{if } i \in [m_{start},
        m_{start}+s*x_{len}] \\
        0, & \text{otherwise} \end{cases}$ \\
        \vspace{0.1cm}
        \textit{\# Generate noisy pattern within input limits}

        $P = \texttt{rand\_tensor}(x_{min}, x_{max}, \text{type}=x.\text{type})$\\
        \KwRet $M, P$
    }
    \vspace{0.1cm}
    \Def{apply\_pattern($\text{input } x$)}{
        $ x^*(i) = 
                \begin{cases} x[i], & M(i) = 1 \\
                            P[i], & M(i) = 0 \end{cases}$ \\
        \KwRet $(x^*, y^*)$
    }
    \vspace{0.1cm}
    
    \Def{\_\_getattr\_\_(dataset attribute $a$)}{
        \textit{\# Mirror all other methods from $\mathcal{D}$} \\
        \KwRet \texttt{getattr}$(\mathcal{D}, a)$
    }
    }
    \end{algorithm}

Algorithm~\ref{alg:poison} demonstrates how to create the mask
and the pattern for a generic dataset $\mathcal{D}$ with tuples
$(x_i, y_i), \forall i \in I$ indices.
We use 1D inputs
for brevity.  Multidimensional inputs can be similarly reshaped to 1D
to build the mask.  This approach can be used to poison any dataset by
(a) identifying indices $I^* \subset I$ to poison, and (2) overwriting
the $\_\_\texttt{get}\_\_(i)$ method to invoke $\texttt{apply\_pattern}$
for inputs $i \in I^*$.  The wrapper is transparent to all other methods
and attributes by forwarding calls to the original dataset.

Similarly, when validating a model $\theta$, in addition to
computing the main-task accuracy $\textbf{A}(\mathcal{D}_{val},
\theta)$, \method uses the same wrapper but sets the poisoning
percentage $p_{\star}=1$ to measure accuracy of the primitive sub-task
$\textbf{A}(\mathcal{D}^{p_\star=1}_{val}, \theta)$.  This design allows
\method to inject any type of backdoor and measure its efficacy.

\section{Preventing Accidental Success} 
\label{sec:preventing_success}

When picking the pattern and the label for a primitive backdoor attack,
it is important to avoid cases when the backdoor label would already
be a highly likely candidate because this can skew measurements of
attack efficacy.  For example, attacking a sentiment classification
task with the backdoor trigger ``awesome'' and backdoor label
``positive'' will show seemingly high efficacy since the backdoor task
is correlated with the main task.  This effect is described as natural
backdoors~\cite{wengerfinding, wang2022training, tao2022backdoor}, i.e.,
naturally occurring features of the dataset that have a strong signal for
particular classes.  These attacks are only feasible if the attacker has
access to the dataset and/or model and are outside the threat model of
this paper.  Therefore, after picking the trigger and backdoor label,
we can simply test the non-poisoned model to verify that it does not
already exhibit high backdoor accuracy.

Another reason to test the non-poisoned model is adversarial
patches~\cite{brown2017adversarial}.  This inference-time attack covers
$5-10\%$ of the input and causes the model to switch its output without
any data poisoning\textemdash but the model does not learn a backdoor
task.  Since we create patterns randomly, we may generate an adversarial
patch by accident, thus it is important to check that the model does not
already (i.e., without poisoning) associate the pattern with some label.

We should also assume that an unknown fraction of the training dataset
$\mathcal{D}$ may already be poisoned with an unknown number of backdoors.
Therefore, our injection of the primitive sub-task may accidentally
select already-poisoned inputs, or else cover backdoor triggers.  We can
similarly test the accuracy of the non-poisoned model to avoid collisions.

\end{document}